\documentclass{article}
\usepackage{arxiv}
\usepackage[utf8]{inputenc}
\usepackage[T1]{fontenc}
\usepackage{hyperref}
\usepackage{url}
\usepackage{graphicx}
\usepackage{booktabs}
\usepackage{longtable}
\usepackage{array}
\usepackage{calc}
\usepackage{multirow}
\usepackage{amsmath,amssymb,amsfonts}
\usepackage{microtype}
\usepackage{enumitem}
\usepackage{ragged2e}
\usepackage{cite}
\usepackage{xcolor}
\usepackage{float}
\usepackage{placeins}
\newcommand{\real}[1]{#1}
\hypersetup{colorlinks=true,linkcolor=blue,citecolor=blue,urlcolor=blue}

\title{Learning continuous reaction paths for transition-state prediction}
\author{Yexiang Yang, Linlin Zhong\thanks{Corresponding author: linlin@seu.edu.cn} \\
School of Electrical Engineering, Southeast University, Nanjing, China}
\date{September 20, 2026}
\renewcommand{\headeright}{}
\renewcommand{\undertitle}{A Preprint}

\begin{document}
\maketitle

\begin{abstract}
Transition states are defined by reaction pathways, yet most
machine-learning methods predict them as isolated geometries. We
introduce MARC-TS, a two-stage framework that learns a continuous,
endpoint-conditioned path, queries it at any resolution and uses local
path context to refine a transition-state candidate. We construct
T1x-IRC-8K, a dataset of 8,209 reactions and 1,088,725 path-resolved
geometries. On held-out reactions, the path model reduced complete-path
error by 48.4\% relative to endpoint interpolation, and the localizer
achieved a mean aligned structural error of 0.127 \AA{}. Quantum-chemical
optimization and vibrational analysis yielded 405 frequency-confirmed
first-order saddle-point candidates from 410 predictions. In a
100-reaction nudged elastic band comparison, learned-path initialization
reached a joint geometry-and-force target for 66\% of reactions,
compared with 12\% for geometric interpolation after 100 optimizer
steps. By treating the path as a reusable representation rather than an
auxiliary output, MARC-TS connects transition-state prediction,
mechanistic interpretation and quantum-chemical refinement.
\end{abstract}

\section{Introduction}
Chemical reactions pass through a transition state: a fleeting,
high-energy arrangement of atoms that helps determine reaction rates and
selectivity. Locating these structures is central to mechanistic
chemistry {[}1{]}, catalyst design {[}2{]} and kinetic modelling
{[}3{]}, but remains computationally demanding. A transition state is a
first-order saddle point on the potential-energy surface, with energy
rising along the reaction direction and falling along all orthogonal
directions. Conventional workflows refine an initial guess with
quantum-chemical forces and test the result by vibrational analysis and
reaction-path calculations {[}4--6{]}. Because refinement is local, a
poor starting geometry can converge slowly, reach another saddle region
or fail.

Chain-of-states methods make the reaction route explicit by optimizing
structures between known endpoints {[}7--10{]}. The nudged elastic band
(NEB) method represents the route as an ordered chain of molecular
geometries, or images, and relaxes it using forces resolved along and
across the path. Its cost and outcome depend strongly on initialization.
Cartesian linear interpolation and image-dependent pair-potential (IDPP)
interpolation are inexpensive, but infer the interior from endpoint
geometry alone {[}11{]}. They can compress concerted motion, undersample
the transition region or require extensive force-driven reorganization.
A reaction-specific path prior could provide a better description of the
transformation and a more effective starting point for quantum-chemical
refinement.

Machine learning offers a complementary route. Reaction datasets and 3D
equivariant neural networks can rapidly generate transition-state
candidates from molecular endpoints {[}12,14--17{]}. Coordinate
regression {[}18,19{]}, diffusion {[}20,21{]}, optimal transport
{[}22{]} and flow matching {[}23--25{]} have reduced reliance on
hand-built guesses, while reactive machine-learning potentials can
accelerate subsequent force evaluations {[}26{]}. Most approaches,
however, still predict an isolated geometry. This formulation ignores
information in the physical definition of a transition state: its
position on a path connecting molecular basins. Even an accurate point
prediction neither reveals the concerted motion that produced it nor
supplies neighbouring structures for mechanistic interpretation, failure
diagnosis or path-based optimization.

We instead learn the reaction path first and localize the saddle region
from it. The model represents molecular geometry as a continuous
function of reaction progress conditioned on the endpoints, rather than
as a sequence of fixed length. This separates the learned representation
from its discretization: one inference can provide a candidate
structure, a local window, an image chain at chosen resolution or a
dense trajectory. Recent generative models can also produce pathways
{[}24,25{]}; our formulation learns a deterministic field over a
physically normalized coordinate and reuses it both for local
transition-state prediction and for force-based path optimization.

We call this framework MARC-TS, for mass-weighted arc-length reaction
coordinates for transition states. It decomposes transition-state
prediction into two linked tasks: learning the global reaction path and
locating the transition state within it. MARC-Field addresses the first
task: it is an endpoint-conditioned, rotation- and
translation-equivariant neural field that returns a geometry at any
requested path coordinate. The second component, MARC-Loc, reads a
five-frame segment of the predicted path and refines its central
structure into a transition-state candidate. The path is therefore a
shared representation that can be evaluated globally, examined locally
or discretized for chain-of-states calculations. An endpoint-union graph
links atoms that are bonded or spatially close in either endpoint,
including regions where bonds form or break. The individual components,
including SE(3)-equivariant message passing, latent interpolation, and
iterative geometry updates, are established. The contribution is their
organization around a path-based prediction task, a mass-weighted
progress coordinate, a field-to-localizer design and the path-resolved
supervision required to train it.

Existing endpoint--transition-state collections do not provide path
supervision, so we constructed T1x-IRC-8K from Transition1x {[}13{]}. It
contains 8,209 endpoint-consistent reactions and 1,088,725 intrinsic
reaction-coordinate (IRC) geometries, preserving how structures evolve
between molecular basins. We ask whether MARC-Field learns interior
motion beyond endpoint interpolation; whether MARC-Loc benefits from
local path context; whether its predictions enter plausible saddle
regions under density-functional-theory (DFT) refinement; whether
geometry-only paths preserve energy-profile structure; and whether the
representation remains useful with independently optimized endpoints and
for NEB initialization. We distinguish geometric agreement, local saddle
character and endpoint connectivity because none implies the others.
More broadly, the study tests a machine-learning principle relevant to
other pathway-dependent problems: a target defined by a trajectory may
be better represented by learning the trajectory than by predicting one
state.

\section{Results}\label{results}

\subsection{Path-grounded MARC-TS framework and T1x-IRC-8K
dataset}\label{path-grounded-marc-ts-framework-and-t1x-irc-8k-dataset}

Figure 1a contrasts direct transition-state prediction with the
two-stage MARC-TS framework. For a reaction containing \emph{N}
consistently mapped atoms, let the reactant and product Cartesian
coordinates define the two boundary geometries, and let \textbf{Z}
denote the atomic numbers. MARC-Field represents the path as an
endpoint-conditioned neural field that is equivariant to rigid rotations
and translations:

\begin{equation*}
\hat{\mathbf X}_{\mathrm{pred}}(u)=F_{\theta}(\mathbf X_{\mathrm R},\mathbf X_{\mathrm P},\mathbf Z,u),\qquad u\in[0,1].
\end{equation*}

The learned parameters are denoted by \emph{$\theta$}, and the output is the
predicted molecular geometry at reaction-progress coordinate \emph{u}.
We define \emph{u} by normalized cumulative mass-weighted arc length,
with \emph{u} = 0 at the reactant and \emph{u} = 1 at the product. Mass
weighting gives greater influence to the same displacement of a heavier
atom than to that of hydrogen, while normalization places trajectories
of different total length on a common interval. Querying different
values of \emph{u} allows one trained model to generate paths at any
desired resolution.

As shown in Fig. 1b,c, a shared encoder maps the two endpoint geometries
into latent representations, which are mixed at the requested value of
\emph{u}. An SE(3)-equivariant decoder converts the resulting
representation into Cartesian coordinates. MARC-Loc then queries five
structures at \emph{u} = \{0.40, 0.45, 0.50, 0.55, 0.60\} and refines
the central structure. This window describes how the predicted path
approaches and leaves the transition region. First- and
second-difference features summarize, respectively, the local direction
of motion and how that direction changes. The reference transition-state
geometry and its coordinate are used only for supervision and post hoc
analysis; neither is supplied to MARC-Field or MARC-Loc at inference.

To provide the path-resolved supervision required by MARC-TS, we
constructed T1x-IRC-8K by tracing the intrinsic reaction coordinate
(IRC) in both directions from each of the 10,073 reference transition
states in Transition1x {[}13{]}. Each IRC follows the local
steepest-descent path from a transition state towards its connected
reactant and product basins in mass-weighted coordinates. Both
directions terminated normally for 8,939 trajectories; 730 were excluded
because at least one endpoint could not be matched to the reactant or
product assigned in Transition1x. The curated dataset contains 8,209
endpoint-consistent reactions and 1,088,725 geometries from neutral C,
H, N and O molecules, with a median of 14 atoms. Resolution varies
substantially: trajectories contain a median of 122 frames and a 90th
percentile of 211 frames (Fig. 1d). The transition state is near, but
not fixed at, the midpoint; the median reference coordinate is \emph{u}
= 0.502 (Fig. 1e). This variation favours a coordinate-conditioned field
over a fixed midpoint or fixed-length output.

We used the final sampled IRC structures as model boundaries rather than
independently optimized minima. This keeps both endpoints, all
intermediate structures, and the reference transition state on one
continuous trajectory, so each query coordinate refers to a consistent
path. Across 16,418 endpoints, the median all-atom root-mean-square
deviation (RMSD), after optimal rigid alignment, between an IRC endpoint
and its optimized Transition1x counterpart was 0.053 \AA{}; excluding
hydrogen reduced the median to 0.022 \AA{} (Supplementary Fig. S1). The
shift is small for most reactions but systematic, so we assessed its
effect separately in the endpoint-domain adaptation experiment below.

All analyses used a predefined split of 7,409 training, 390 validation
and 410 held-out test reactions. To generate MARC-Loc training inputs
without in-sample MARC-Field predictions, we used five-fold
cross-fitting: each training reaction was processed by a MARC-Field
model that had not seen it during training. A final MARC-Field model
trained on all 7,409 training reactions generated paths for the
validation and held-out sets. Validation data were used to select
checkpoints. The held-out set was evaluated only after the models,
thresholds and analysis procedures had been fixed (Supplementary Fig.
S2).

\begin{figure}[H]
\centering
\includegraphics[width=\linewidth,height=0.62\textheight,keepaspectratio]{figures/Fig1.png}
\par\smallskip
\begin{minipage}{0.98\linewidth}\small
\textbf{Fig. 1 Path-grounded MARC-TS framework and the
T1x-IRC-8K dataset. a}, Conceptual comparison of direct pointwise
transition-state prediction and path-grounded prediction. \textbf{b},
MARC-Field encodes the reactant and product endpoints and interpolates
their latent representations at a requested mass-weighted coordinate
\emph{u}. An SE(3)-equivariant decoder generates the corresponding
molecular geometry. \textbf{c}, MARC-Loc refines the central prediction
using a five-frame window generated at \emph{u} = \{0.40, 0.45, 0.50,
0.55, 0.60\}. \textbf{d}, Native IRC resolution as a function of
molecular size. Points are reactions, the line is the median and shading
is the interquartile range. \textbf{e}, Distribution of reference
transition-state positions along \emph{u}; the dashed line is the
median. Dataset statistics use all 8,209 reactions.
\end{minipage}
\end{figure}

\subsection{MARC-Field captures reaction-specific path
geometry}\label{marc-field-captures-reaction-specific-path-geometry}

We first asked whether MARC-Field learns reaction-specific interiors
rather than a smoothed endpoint interpolation. Accuracy was measured by
all-atom RMSD after optimal rigid alignment, which removes overall
translation and rotation and isolates internal geometry. For scale, an
RMSD of 0.1 \AA{} is roughly one tenth of a common covalent-bond length.
Each reaction was evaluated at 11 equally spaced coordinates from
\emph{u} = 0 to 1. Cross-fitting supplied out-of-fold training
predictions, and the final frozen model supplied validation and held-out
predictions. Linear interpolation used the same aligned endpoints and
coordinates, so the methods differed only in how they inferred the path
interior.

Across 8,209 reactions, MARC-Field reduced mean complete-path RMSD from
0.196 to 0.101 \AA{}, a 48.4\% decrease, and improved 96.7\% of reactions
(Fig. 2a; two-sided paired Wilcoxon \emph{P} = 4.5 $\times$
10\textsuperscript{-7}). The gain was therefore broadly distributed
rather than driven by a few outliers. Mean and 90th-percentile errors
remained below interpolation throughout the transition region (Fig. 2b),
and the advantage persisted across complete, interior, central, and
transition-state-associated coordinates (Fig. 2c).
Predicted-to-reference mass-weighted path-length ratios clustered near
one, whereas interpolation systematically shortened the paths (Fig. 2d).
Because path length was not optimized directly, its preservation
suggests that MARC-Field learned aspects of the route between endpoint
basins rather than only isolated queried structures.

Upper-tail and path-length diagnostics are reported in Supplementary
Fig. S3. Error increased with atom count, reference path length and
native IRC resolution, but MARC-Field retained an advantage in every
quartile and across their joint grid (Supplementary Fig. S4). High-error
reactions generally involved more extensive collective rearrangements,
although MARC-Field usually remained closer to the reference in the
transition region. Together, these analyses support robustness across
the reaction complexity represented in T1x-IRC-8K; transfer to chemical
classes outside this dataset remains untested.

\begin{figure}[H]
\centering
\includegraphics[width=\linewidth,height=0.62\textheight,keepaspectratio]{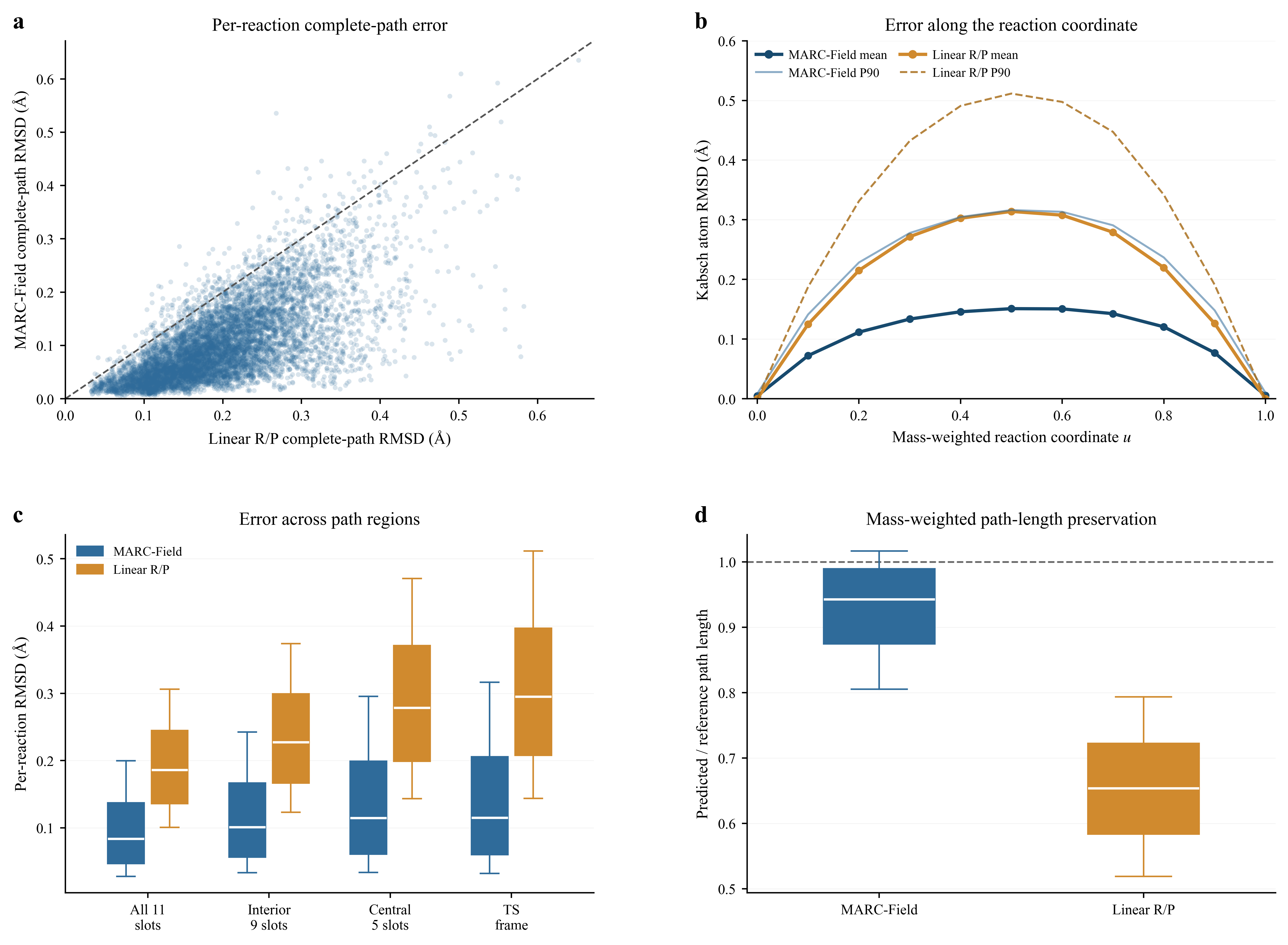}
\par\smallskip
\begin{minipage}{0.98\linewidth}\small
\textbf{Fig. 2 MARC-Field captures reaction-path geometry
beyond linear endpoint interpolation.} \textbf{a}, Per-reaction
complete-path RMSD for MARC-Field and linear reactant/product
interpolation; the dashed diagonal indicates equal error. \textbf{b},
Mean and 90th-percentile all-atom RMSD along the normalized
mass-weighted coordinate. \textbf{c}, Error distributions over all 11
path positions, the 9 interior positions, the 5 central evaluation
positions and the position associated with the reference transition
state. \textbf{d}, Predicted-to-reference mass-weighted path-length
ratios; the dashed line indicates exact preservation. Box plots show
medians, interquartile ranges and 10th--90th-percentile whiskers. All
panels summarize 8,209 reactions. The central five-position evaluation
subset \emph{u} = \{0.30, 0.40, 0.50, 0.60, 0.70\} differs from the
five-frame MARC-Loc input window.
\end{minipage}
\end{figure}

\subsection{Local path context improves transition-state
localization}\label{local-path-context-improves-transition-state-localization}

The central MARC-Field prediction at \emph{u} = 0.50 served as the
anchor and had a mean all-atom RMSD of 0.184 \AA{} on the 410-reaction
held-out set. MARC-Loc improved 83.7\% of these structures and reduced
the mean error by 31.2\%, to 0.127 \AA{} (Fig. 3a). Improvements appeared in
every quartile of anchor error, with the largest absolute corrections in
the most difficult quartile. A fixed midpoint keeps the inference
procedure simple but cannot account for reactions whose transition state
is displaced along the path. Supplementary Fig. S5a--c quantifies
sensitivity to the five candidate anchor positions and the penalty of
fixing \emph{u} = 0.50.

To isolate the value of path context, we compared MARC-Loc with
DirectTS, an endpoint-only model using a matched backbone and the same
data split. DirectTS achieved mean and median RMSDs of 0.142 and 0.115
\AA{}, whereas MARC-Loc reduced these values by 10.9\% and 21.8\%,
respectively (Fig. 3b). We also used a post hoc window ORACLE: a
reference-informed diagnostic, unavailable in deployment, that selects
the best of the five input structures. MARC-Loc outperformed every input
structure for 58.3\% of reactions (Supplementary Fig. S5c). The
localizer therefore does not merely select a favourable frame; it
combines the local motion encoded by the window to generate a new
geometry.

Similarity to a reference structure does not establish transition-state
character. We therefore used every held-out MARC-Loc prediction to
initialize an independent quantum-chemical saddle-point search. This
tests whether the prediction lies within the convergence region of a
first-order saddle point; subsequent vibrational analysis tests whether
the optimized structure has exactly one direction of negative curvature.
The primary protocol converged for 367 reactions, and a predefined
restart from the original prediction recovered 40 more. Overall, 407 of
410 optimizations completed successfully (99.3\%), and 405 of the
resulting structures had exactly one significant imaginary vibrational
mode (Fig. 3c).

We further compared each unstable mode with the local tangent of the
reference IRC and with the atom pairs whose bonding changes between the
endpoints. Among the 405 frequency-confirmed candidates, 89.1\% had a
tangent-overlap score of at least 0.99 and 71.9\% had a reactive-pair
sign-agreement score of at least 0.90 (Fig. 3d). These results support
local first-order saddle character and consistency of the unstable
direction. They do not prove that the optimized structure connects the
intended reactant and product; that stronger claim requires
bidirectional IRC calculations, which were not performed here.

Table 1 compares MARC-TS with local evaluations of React-OT {[}22{]},
LearnTS {[}19{]}, OA-ReactDiff {[}20{]}, GoFlow {[}23{]} and TSDiff
{[}21{]}, but the values are not fully comparable. Methods given 3D
endpoints and those given only a 2D reaction graph solve different
tasks. Atom rematching, reflection and aggregation also differ.
Moreover, local evaluations do not reproduce every published value;
React-OT, for example, reported mean and median RMSDs of 0.103 and 0.053
\AA{} on Transition1x after its published initialization and fine-tuning
procedure, versus 0.160 and 0.126 \AA{} here. We therefore make no universal
state-of-the-art claim. The controlled comparison supporting our
conclusion is MARC-Loc versus DirectTS, which shares the data, endpoint
information and backbone while removing path context. Supplementary
Table S1 documents all protocol differences.

\begin{table}[H]
\small
\textbf{Table 1 Transition-state geometry comparison on 410
held-out reactions. Values are mean and median all-atom RMSDs (\AA{}); lower is better.} Only evaluations performed in this work are
included. Reactant/product denotes endpoint input and EMA denotes
exponential moving average. Methods conditioned on 3D endpoints and
those conditioned on 2D reaction graphs solve different tasks and should
not be compared directly. Alignment, atom-rematching, reflection, and
aggregation conventions are given in Methods and Supplementary Table S1.

\medskip
\centering
\begin{tabular}{@{}>{\raggedright\arraybackslash}p{0.22\linewidth} >{\raggedright\arraybackslash}p{0.47\linewidth} >{\raggedright\arraybackslash}p{0.25\linewidth}@{}}
\toprule
\textbf{Method} & \textbf{Input and local checkpoint} & \textbf{Mean / median RMSD (\AA{})} \\
\midrule
\textbf{MARC-TS} & \textbf{3D R/P plus five-frame path context} & \textbf{0.1269 / 0.0902} \\
DirectTS & 3D R/P; endpoint-only matched control & 0.1424 / 0.1153 \\
React-OT {[}22{]} & 3D R/P; local 1,000-epoch reproduction & 0.1604 / 0.1256 \\
LearnTS {[}19{]} & 3D R/P plus interpolation; epoch 1453 & 0.1708 / 0.1410 \\
OA-ReactDiff {[}20{]} & 3D R/P; epoch 2,000 EMA & 0.1891 / 0.0965 \\
GoFlow {[}23{]} & 2D reaction graph; epoch 650 & 0.2957 / 0.2104 \\
TSDiff {[}21{]} & 2D reaction graph; iteration 67,000 & 0.5633 / 0.6040 \\
\bottomrule
\end{tabular}
\end{table}

The primary models use terminal IRC frames as boundaries, keeping every
structure on one trajectory. Practical applications instead start from
independently optimized reactants and products, a shifted input domain.
We therefore fine-tuned MARC-Loc on windows generated from optimized
endpoints. On the held-out set, the adapted model achieved mean and
median RMSDs of 0.193 and 0.136 \AA{} (Supplementary Fig. S6). Fine-tuning
recovered useful accuracy but did not close the gap, and its mean error
exceeded that of React-OT under the local protocol in Table 1. Rankings
from native IRC endpoints should therefore not be extrapolated to
deployment. This mismatch is the principal practical limitation and
motivates training paths computed directly between optimized minima.

\begin{figure}[H]
\centering
\includegraphics[width=\linewidth,height=0.62\textheight,keepaspectratio]{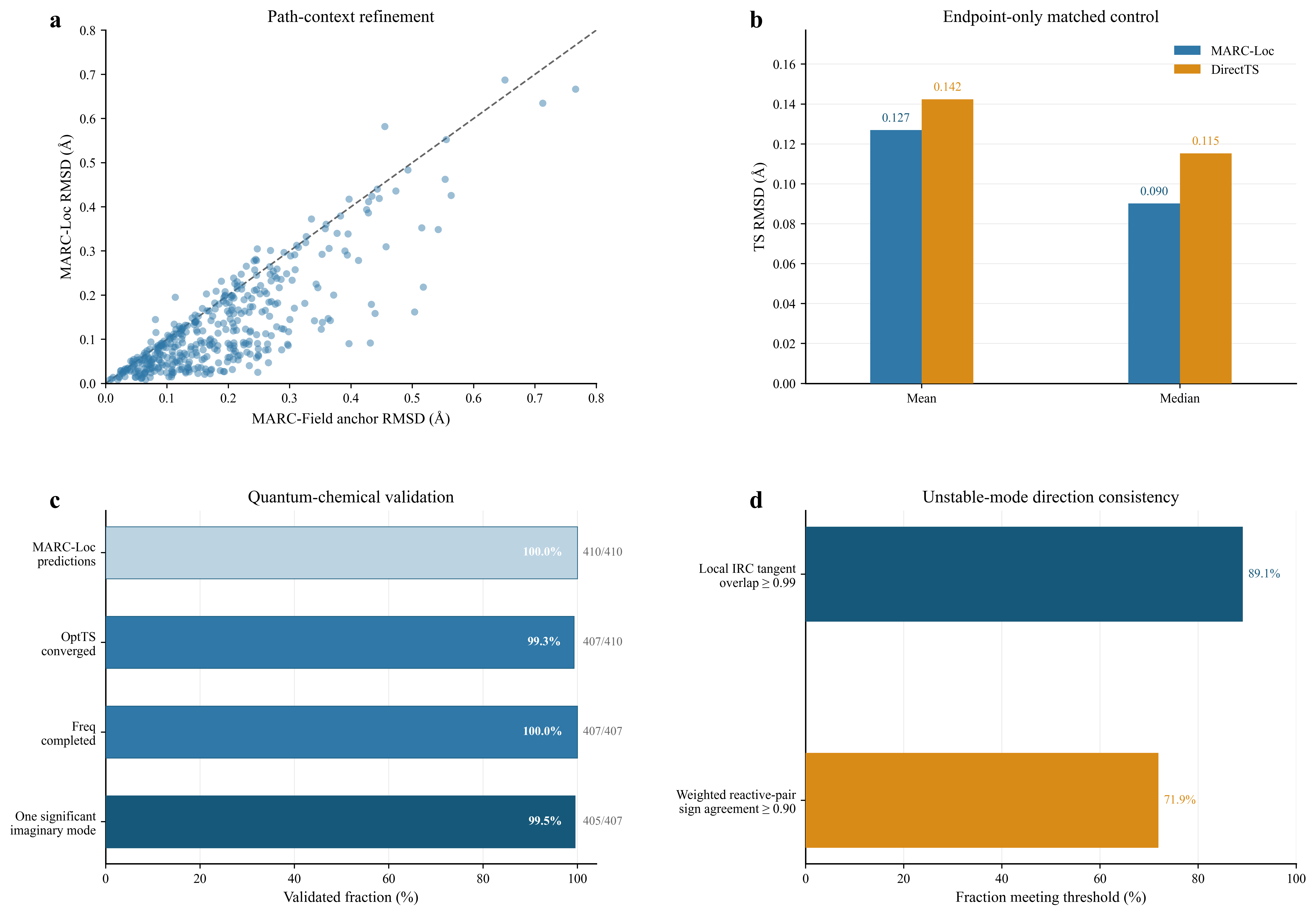}
\par\smallskip
\begin{minipage}{0.98\linewidth}\small
\textbf{Fig. 3} \textbf{Local path context improves
transition-state geometry and supports quantum-chemical validation. a},
Per-reaction MARC-Loc RMSD versus the central MARC-Field anchor for 410
held-out reactions; the dashed diagonal indicates equality. \textbf{b},
Mean and median RMSDs for MARC-Loc and the endpoint-only DirectTS
control. \textbf{c}, Fractions completing Gaussian transition-state
optimization and frequency analysis, and the fraction with exactly one
significant imaginary mode. \textbf{d}, Consistency of the unstable mode
with the local IRC tangent and endpoint-changing atom pairs. Of 405
frequency-confirmed candidates, 89.1\% had tangent overlap $\geq$ 0.99 and
71.9\% had reactive-pair sign agreement $\geq$ 0.90. A significant imaginary
mode was \emph{$\nu$} \textless{} -20 cm\textsuperscript{-1}. These
diagnostics support local saddle character and direction consistency but
do not establish endpoint connectivity.
\end{minipage}
\end{figure}

\subsection{Geometry-only learning preserves coarse energy-landscape
structure}\label{geometry-only-learning-preserves-coarse-energy-landscape-structure}

MARC-Field is trained only on geometry, without energies, forces or
gradients. We therefore asked whether generated paths retain features of
independently evaluated DFT energy profiles. MARC-Field was queried at
every stored reference coordinate for all 410 held-out reactions, and
all 53,888 single-point calculations at $\omega$B97X/6-31G(d) terminated
normally. Reference and predicted profiles were shifted independently to
zero minimum, removing absolute offsets and focusing on relative
variation along the path. At the population level, the predicted median
followed the location and broad shape of the reference high-energy
region (Fig. 4a), with a median per-reaction profile root-mean-square
error (RMSE) of 6.11 kcal mol\textsuperscript{-1} (Fig. 4b).

The predictions also retained several interpretable profile features.
Energy spans and endpoint-referenced forward and reverse barrier-like
descriptors tracked their reference values (Supplementary Fig. S7). The
median absolute displacement of the maximum-energy coordinate was about
0.025 on the normalized path, and the predicted maximum lay within 0.10
of the reference position for 91.0\% of reactions (Fig. 4c). Reactions
with larger geometric path errors also tended to have larger profile
RMSEs and energy-span errors, suggesting that path RMSD may flag cases
in which the energetic picture is less reliable.

The magnitude of this error sets a clear limit: 6.11 kcal
mol\textsuperscript{-1} (about 0.26 eV) is too large for quantitative
thermochemistry. The profiles are not activation-barrier predictions.
Instead, they show that geometry-only learning often preserves coarse
profile shape and peak location. These single-point energies at fixed
geometries are not force-converged minimum-energy paths, activation
energies or Hessian-confirmed transition states. Supplementary Figs. S8
and S9 show examples across the distribution and separate errors in
scale, shape and peak position.

Figures 4d--f show representative reactions spanning the observed range,
from close agreement between path geometry and energy profiles to cases
in which that correspondence breaks down. Reaction 3810 is a favourable
case in which path geometry, peak position and both predicted
transition-state geometries closely match their references. Reaction
2636 is typical: the main motion is preserved, the energy profile
remains displaced and local refinement partly corrects the
transition-state geometry. Reaction 571 is an atypical failure with a
carbene-like product centre and unusual product-side electronic
structure. It should not be interpreted as a general broadening of
transition regions. These cases show when geometric fidelity preserves
coarse energetic organization and when that relationship fails.

\begin{figure}[H]
\centering
\includegraphics[width=\linewidth,height=0.62\textheight,keepaspectratio]{figures/Fig4.png}
\par\smallskip
\begin{minipage}{0.98\linewidth}\small
\textbf{Fig. 4 MARC-Field trajectories retain coarse features
of the DFT energy landscape. a}, Aggregate reference IRC and MARC-Field
single-point energy profiles along \emph{u}. Lines show medians, shading
shows interquartile ranges and the vertical band marks \emph{u} =
0.40--0.60. \textbf{b}, Cumulative distribution of per-reaction
relative-energy profile RMSE. \textbf{c}, Reference versus predicted
maximum-energy coordinate. \textbf{d--f}, Representative favourable
(reaction 3810), typical (reaction 2636) and failure (reaction 571)
cases, combining path geometries, energy profiles and reference,
MARC-Field and MARC-Loc transition-state geometries.
\end{minipage}
\end{figure}

\subsection{MARC-Field improves NEB
initialization}\label{marc-field-improves-neb-initialization}

We next tested MARC-Field as an initializer for force-based path
optimization. This benchmark compares complete workflows rather than the
neural geometry alone. MARC-Field used 11 DFT single-point calculations
to centre the energy peak, followed by 100 climbing-image NEB steps;
IDPP used no scan and 30 ordinary plus 70 climbing-image steps. Thus,
only post-construction optimizer steps---not total electronic-structure
cost or phase schedule---were matched. Across 100 paired reactions,
final median centre-image RMSD was 0.041 \AA{} for MARC-Field and 0.120 \AA{}
for IDPP. Median maximum atomic force fell from 0.826 to 0.016 eV
\AA{}\textsuperscript{-1}, and median maximum projected NEB force from 0.084
to 0.042 eV \AA{}\textsuperscript{-1} (Fig. 5a--c).

Before optimization, about 69\% of MARC-Field bands already met the
centre-image geometry target of RMSD $\leq$ 0.10 \AA{}. After 100 steps, 87\% of
MARC-Field bands and 60\% of IDPP bands met that target (Fig. 5d). The
force target, a maximum centre-image atomic force of $\leq$ 0.05 eV
\AA{}\textsuperscript{-1}, was reached by 90\% and 47\%, respectively (Fig.
5e). A joint target requiring both conditions, fixed before evaluation,
was reached by 66\% of MARC-Field bands and 12\% of IDPP bands (Fig.
5f).

The final highest-energy image need not be the centre image. Its median
geometric error was 0.033 \AA{} for MARC-Field and 0.058 \AA{} for IDPP; median
relative-energy errors were 0.021 and 0.146 kcal
mol\textsuperscript{-1}. MARC-Field had the lower error in 72\% of
paired reactions for both metrics (Supplementary Fig. S10a--c).
Separately, the complete MARC-Field-to-MARC-Loc pipeline required a
median of 122.9 ms per reaction on one NVIDIA RTX A6000 GPU, with
approximately linear scaling in requested path coordinates
(Supplementary Fig. S11). Neural inference is therefore inexpensive
relative to the electronic-structure calculations that dominate
end-to-end cost.

\begin{figure}[H]
\centering
\includegraphics[width=\linewidth,height=0.62\textheight,keepaspectratio]{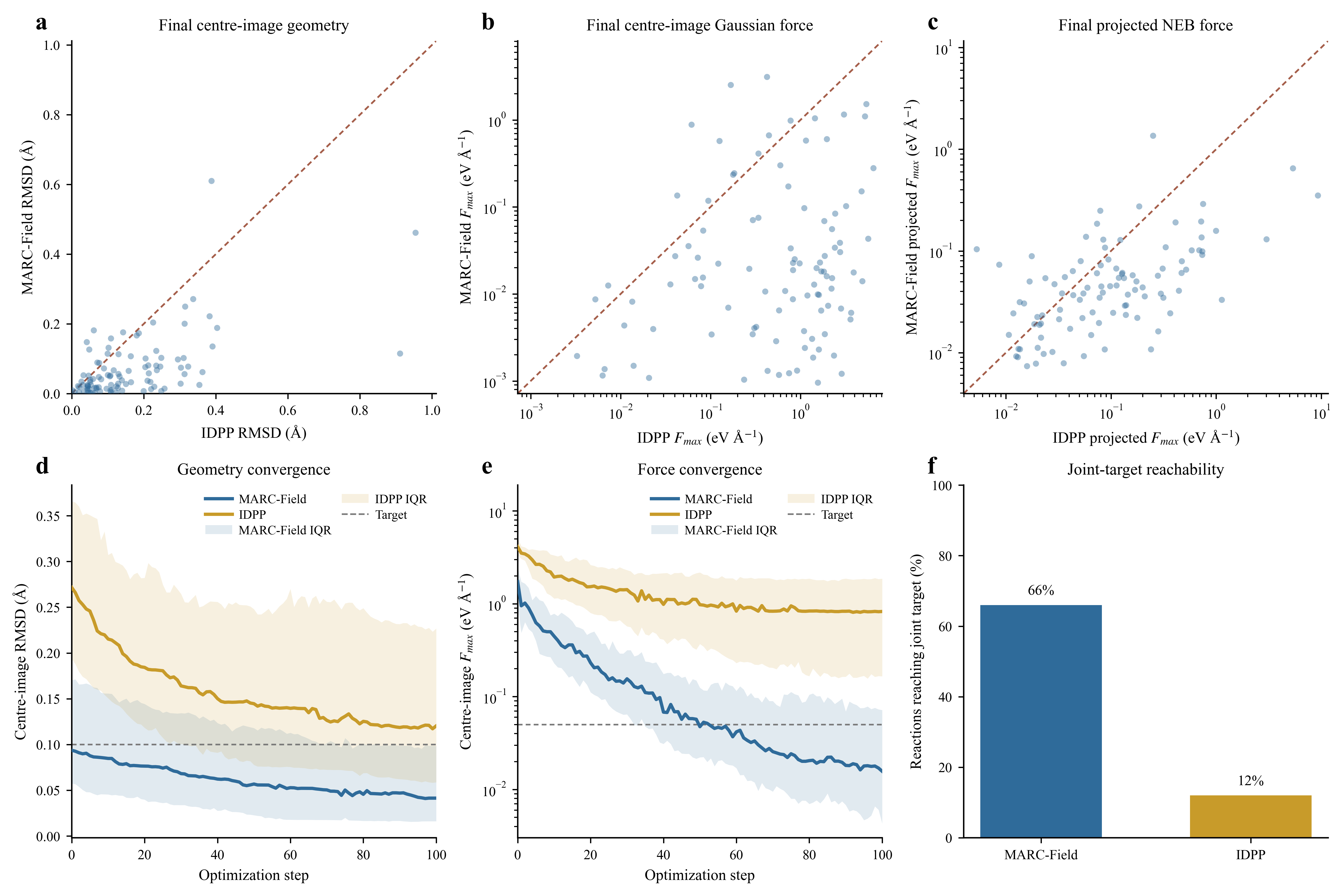}
\par\smallskip
\begin{minipage}{0.98\linewidth}\small
Fig. 5 MARC-Field initialization improves NEB outcomes within
the predefined workflow. \textbf{a--c}, Final centre-image
transition-state RMSD, maximum atomic force on the centre image and
maximum projected NEB force for 100 paired MARC-Field and IDPP
calculations. Dashed diagonals indicate equal performance. \textbf{d,e},
Evolution of centre-image RMSD and maximum atomic force over 100
optimizer steps. Lines show medians, shading shows interquartile ranges
and dashed horizontal lines mark the predefined targets. \textbf{f},
Fraction of reactions reaching both targets. The workflows used the same
endpoints, 11 images, electronic-structure level and number of
post-construction optimizer steps, but not the same total
electronic-structure cost or phase schedule: MARC-Field used an 11-point
DFT peak-centring scan and 100 climbing-image steps, whereas IDPP used
no scan and 30 ordinary plus 70 climbing-image steps. The joint target
was RMSD $\leq$ 0.10 \AA{} and maximum centre-image atomic force $\leq$ 0.05 eV
\AA{}\textsuperscript{-1}. Unreached reactions remained in the denominator.
\end{minipage}
\end{figure}

\section{Discussion}\label{discussion}

MARC-TS recasts transition-state prediction as learning a continuous
path and then extracting the local structure of interest. This is more
than predicting a fixed collection of additional geometries: a path
records how a system enters and leaves the transition region and can be
sampled at the resolution required downstream. One field can therefore
support full-path generation, local transition-state prediction,
visualization and chain-of-states initialization. The central
contribution is this reusable representation, enabled by a mass-weighted
coordinate, a field-and-localizer architecture and path-resolved
supervision.

The matched controls isolate what the path contributes. Endpoints
already contain substantial information, as the competitive DirectTS
results show. Even so, MARC-Loc improves on that endpoint-only model
under a common backbone and data split. It also generates a structure
better than every member of its input window for most reactions. Local
path context therefore provides information beyond selecting a
favourable predicted frame.

T1x-IRC-8K enables this formulation by retaining a curated IRC
trajectory rather than only reactant, transition-state and product
structures. It supports complete-path error, path-length preservation,
transition-position and energy-profile analyses that endpoint triplets
cannot. The normalized coordinate separates the representation from
source sampling density, allowing new query positions without changing
the learning problem. Public release of the trajectories, split and
evaluation tables is therefore essential for reproducible path-level
comparison.

Quantum-chemical analyses clarify how MARC-TS complements established
calculations. MARC-TS supplies candidate geometries and image strings,
while optimization, vibrational analysis and unstable-mode agreement
provide local tests of saddle-point character. These analyses show that
MARC-Loc often enters a locally appropriate saddle region, and
independent single-point profiles indicate that geometry-only paths can
preserve coarse energetic organization. In the NEB benchmark, MARC-Field
initialization improved geometry and force outcomes within the
predefined 100-step workflow. Because that workflow also used an
11-point DFT scan to centre the energy peak and a different optimization
schedule, the result supports initialization utility; equal-total-cost
acceleration remains untested.

The endpoint-domain experiment identifies both the adaptability of the
representation and the next step towards deployment. Native IRC
endpoints provide internally consistent supervision, whereas
independently optimized endpoints define the input domain used in
practice. Fine-tuning recovered useful accuracy on this shifted domain,
showing that MARC-Loc can adapt beyond the native trajectory setting.
Residual upper-tail error and changed method rankings nevertheless show
that endpoint-domain shift remains a material constraint. Training on
paths computed directly between independently optimized minima would
align supervision more closely with deployment and is a clear route for
future improvement.

The scope of the present evidence is set by T1x-IRC-8K and the
evaluation protocol. The dataset contains neutral C, H, N and O
reactions at one DFT level, so charged and open-shell species, heavier
elements, organometallic chemistry and large conformational changes
require separate validation. The model addresses one elementary step and
uses a fixed MARC-Loc window around \emph{u} = 0.50; an adaptive
estimate of the transition coordinate could improve reactions with
displaced saddles. Calibrated uncertainty is another natural extension:
ensemble disagreement or path-consistency measures could identify cases
that merit additional quantum-chemical effort. These considerations
define the next validation and extension steps for the framework.

Taken together, the results support a broader machine-learning design
principle. Many scientific targets, including biomolecular
conformational changes, solid-state diffusion and rare molecular
rearrangements, are defined by trajectories rather than isolated states.
When endpoints are observed but the route is not, an
endpoint-conditioned field over normalized progress can retain
information absent from a single configuration and can be queried at the
resolution required by downstream tasks. A local extraction stage can
then identify the configuration relevant to a particular task.
MARC-Field and MARC-Loc instantiate this principle for chemical
transition states, while electronic-structure optimization and
connectivity analysis provide complementary chemical validation. The
same separation between global path learning and local state extraction
may therefore offer a useful template for other pathway-dependent
scientific problems.

\section{Methods}\label{methods}

\subsection{Dataset construction, path coordinate and predefined
evaluation}\label{dataset-construction-path-coordinate-and-predefined-evaluation}

T1x-IRC-8K was constructed from the 10,073 transition-state structures
in Transition1x {[}13{]}. Bidirectional IRC calculations were run with
Gaussian 16 {[}29{]} at the $\omega$B97X/6-31G(d) level. Both directions
terminated normally for 8,939 trajectories; 1,134 did not complete. Of
the completed trajectories, 730 were excluded because one or both
termini could not be assigned consistently to the designated reactant
and product basins. The retained dataset contains 8,209 reactions and
1,088,725 geometries. Each record stores atomic identities, ordered atom
mapping, Cartesian coordinates, the reference transition-state index,
endpoint identities and available electronic energies, forces and
gradients. The neural models used only atomic identities and geometries.
Supplementary Methods provide the Gaussian route sections, charge and
multiplicity settings, termination criteria and endpoint-assignment
checks.

The final sampled IRC structures, rather than separately optimized
minima, served as model boundaries. This keeps the endpoints,
intermediate geometries and reference transition state on one continuous
trajectory. Path position was defined by cumulative mass-weighted
Cartesian arc length and normalized to \emph{u} $\in$ {[}0,1{]}. The model
was evaluated at each requested value of \emph{u}; it did not substitute
the nearest stored IRC frame. Mass weighting makes an equal displacement
contribute more for a heavy atom than for hydrogen, and normalization
gives reactions with different compositions and total path lengths a
common coordinate interval. Supplementary Methods provide the full
definition and dataset fields.

The deterministic split comprised 7,409 training, 390 validation and 410
held-out reactions, with no reaction shared across subsets. Five-fold
cross-fitting prevented a MARC-Field model from generating MARC-Loc
training inputs for a reaction on which that field had been trained.
Each fold-specific model generated paths only for its excluded fold. A
final MARC-Field trained on all 7,409 training reactions was then frozen
and used for validation and held-out inference. Validation selected
checkpoints; the held-out set was evaluated once after the architecture,
preprocessing, thresholds and analysis procedures had been fixed.

\subsection{MARC-Field architecture and
training}\label{marc-field-architecture-and-training}

MARC-Field represents each reaction with an endpoint-union molecular
graph. Two atoms are connected if they are bonded or spatially close in
either endpoint, allowing messages to pass across regions where bonds
form or break. A shared SE(3)-equivariant encoder {[}14,15,27{]} maps
the reactant and product geometries and atom types to scalar and vector
latent states. At a requested coordinate \emph{u}, these endpoint states
are mixed as follows:

\begin{equation*}
S(u)=(1-u)S_{\mathrm R}+uS_{\mathrm P},\qquad
\mathbf V(u)=(1-u)\mathbf V_{\mathrm R}+u\mathbf V_{\mathrm P}.
\end{equation*}

Here \emph{S}\textsubscript{R} and \emph{S}\textsubscript{P} are the
scalar endpoint states, \emph{V}\textsubscript{R} and
\emph{V}\textsubscript{P} are the corresponding vector states, and
\emph{S}(\emph{u}) and \emph{V}(\emph{u}) are their interpolated values.
The coordinate supplies the mixing weights: \emph{u} = 0 recovers the
reactant state, \emph{u} = 1 the product state and intermediate values
combine the two. The subsequent nonlinear decoder means that linear
mixing in latent space does not constrain the Cartesian path to be
linear.

The decoder starts from Cartesian endpoint interpolation and applies
three geometry-recycling updates. After each update, it recomputes edge
distances, radial features and directions from the current geometry. The
production model has latent dimension 512, eight encoder blocks, eight
decoder blocks, 16 attention heads, 20 radial basis functions, a 10 \AA{}
cutoff and a maximum drop-path rate of 0.20. Supplementary Methods give
the complete operations.

During each training visit to a reaction, one native IRC frame was
sampled uniformly. Its exact mass-weighted coordinate was the query and
its geometry the target. Smooth L1 coordinate losses (\emph{$\beta$} = 0.1)
were applied to all three recycling outputs with weights 0.5, 0.5 and
1.0. No energy, force or gradient labels were used.

Each fold-specific model and the final full-training model were
optimized for at most 600 epochs with batch size 32. We used AdamW, a
maximum learning rate of 4 $\times$ 10\textsuperscript{-4}, weight decay 0.05
and a OneCycle schedule whose increasing phase occupied 30\% of
training. The gradient norm was clipped at 0.5, the exponential moving
average decay was 0.999 and training used 32-bit floating-point
precision.

Validation was performed after epoch 1 and every five epochs thereafter
on the 390-reaction validation set. The selected
exponential-moving-average checkpoint minimized mean-squared coordinate
error across reference frames near \emph{u} = 0.1, 0.3, 0.5, 0.7 and
0.9. Error at the reference transition state was monitored but not used
for selection. Training-loss early stopping was not applied.
Fold-specific checkpoints generated cross-fitted training trajectories;
the final checkpoint generated validation and held-out trajectories.
Supplementary Methods and Supplementary Table S2 provide the full
objectives and configuration.

\subsection{MARC-Loc architecture, training, and endpoint
adaptation}\label{marc-loc-architecture-training-and-endpoint-adaptation}

MARC-Loc receives the two endpoints and five MARC-Field geometries
queried at \emph{u} = \{0.40, 0.45, 0.50, 0.55, 0.60\}. The central
structure is the anchor. Features derived from the remaining structures
describe how the path approaches and leaves that anchor, including
endpoint displacements, left and right summaries, the window mean, a
first difference and a second difference. The first difference is
tangent-like; the second captures local change in that direction. In
total, 24 scalar descriptors and 8 equivariant vector channels are
processed on the endpoint-union graph by eight SE(3)-equivariant
residual blocks {[}14,15,27{]}.

MARC-Loc predicts a bounded local correction to the anchor:

\begin{equation*}
\mathbf X_{\mathrm{TS}}
=\mathbf X_a+\delta_{\max}\tanh\!\left(\frac{\Delta\mathbf X_{\mathrm{raw}}}{\delta_{\max}}\right),
\qquad \delta_{\max}=0.80\,\text{\AA}.
\end{equation*}

Here \emph{\textbf{X}}\textsubscript{a} is the central MARC-Field anchor
and $\Delta$\emph{\textbf{X}}\textsubscript{raw} is the unconstrained
correction predicted by MARC-Loc. A component-wise hyperbolic tangent
bounds each Cartesian displacement by \emph{$\delta$}\textsubscript{max} = 0.80
\AA{} before adding it to the anchor. MARC-Loc therefore performs a local
residual refinement rather than generating an independent structure from
the endpoints.

The production model has a latent dimension of 384 and uses 12 attention
heads, 20 radial basis functions and a 10 \AA{} cutoff. The maximum
drop-path rate is 0.20. Its molecule-balanced training objective is

\begin{equation*}
L_{\mathrm{Loc}}
=L_{\mathrm{atom}}+0.20L_{\mathrm{aux}}+0.01L_{\mathrm{pair}}
+3.0L_{\mathrm{RC\text{-}atom}}+0.5L_{\mathrm{RC\text{-}pair}}.
\end{equation*}

The total objective combines molecule-balanced all-atom coordinate,
auxiliary atom-subset, pair-distance, reaction-centre atom and
changing-pair losses. The displayed coefficients are fixed training
weights, not evaluation parameters. Atom-wise and pair-distance terms
use Smooth L1 losses with \emph{$\beta$} = 0.05. The reaction-centre mask
contains atoms in endpoint-changing bond pairs and one neighbouring
shell from either endpoint, concentrating extra weight on bond formation
and cleavage while retaining the global geometry. Supplementary Methods
define all terms, including diagnostic terms whose production weight is
zero.

MARC-Loc was trained on one fixed, cross-fitted five-frame window per
reaction in each epoch, without coordinate noise or window jitter.
Training used at most 100 epochs, batch sizes of 8 for training and 32
for evaluation, AdamW, a maximum learning rate of 6 $\times$
10\textsuperscript{-5}, a lower reference rate of 5 $\times$
10\textsuperscript{-7}, weight decay 0.005 and a OneCycle schedule with
a 15\% increasing phase. The gradient norm was clipped at 3.0 and the
exponential moving average decay was 0.999. Validation occurred after
epoch 1 and every two epochs thereafter. The selected checkpoint
minimized mean all-atom proper-Kabsch transition-state RMSD on the
validation set. Reference transition-state geometry and coordinate were
used only for supervision and post hoc analysis, never as inference
inputs.

For endpoint-domain adaptation, native IRC boundaries were replaced with
independently optimized Transition1x reactants and products while
retaining the split and cross-fitted provenance. MARC-Field was queried
at the exact five coordinates, without nearest-frame substitution.
MARC-Loc was initialized from the native-endpoint
exponential-moving-average checkpoint; the optimizer, scheduler and
epoch counter were restarted, and fine-tuning used only endpoint-adapted
windows. MARC-Field itself was not fine-tuned. Validation selected epoch
4. This is supervised adaptation to a shifted endpoint domain, not
zero-shot transfer.

\subsection{Geometry evaluation and baseline
protocols}\label{geometry-evaluation-and-baseline-protocols}

Predicted and reference structures were centred at their respective
centroids and aligned using the optimal proper Kabsch rotation
\emph{R}\textsuperscript{$\star$} subject to det(\emph{R}\textsuperscript{$\star$})
= +1 {[}28{]}. The all-atom RMSD was defined as

\begin{equation*}
\operatorname{RMSD}=
\left[\frac{1}{N}\sum_{i=1}^{N}
\left\|R^{\star}(\hat{\mathbf X}_i-\bar{\hat{\mathbf X}})-(\mathbf X_i-\bar{\mathbf X})\right\|_2^2
\right]^{1/2}.
\end{equation*}

Here \emph{N} is the number of atoms, \emph{\textbf{X}\textsubscript{i}}
and $\hat{\mathbf{X}}_i$ are the reference and predicted
coordinates, and their overbars denote centroids.
\emph{R}\textsuperscript{$\star$} is the proper rotation minimizing the summed
squared distance between corresponding atoms. The constraint
det(\emph{R}\textsuperscript{$\star$}) = +1 excludes mirror reflection. The
reported RMSD averages over atoms before taking the square root; it is
not normalized over the 3N Cartesian components.

For generative methods that can permute equivalent atom indices, atoms
of the same element were rematched before alignment. MARC-TS, DirectTS,
LearnTS and TSDiff used proper-Kabsch alignment without reflection,
although TSDiff additionally used same-element rematching. Under their
frozen local protocols, React-OT and GoFlow allowed reflection after
same-element rematching. OA-ReactDiff used its method-specific
rematching procedure, allowed reflection and capped each
reaction\textquotesingle s RMSD at 1 \AA{} before aggregation. Because these
conventions differ, the resulting values are not fully metric-matched.

Path error was summarized within each reaction before cohort
aggregation. Summaries covered all queried coordinates, the interior
coordinates, the five central evaluation coordinates and the coordinate
associated with the reference transition state. DirectTS was the matched
endpoint-only control, using the MARC-Field backbone and the same split
but no intermediate path structures. React-OT {[}22{]}, LearnTS
{[}19{]}, OA-ReactDiff {[}20{]}, GoFlow {[}23{]} and TSDiff {[}21{]}
were evaluated from frozen local checkpoints. TSDiff used one sample per
reaction and 5,000 Langevin steps from iteration 67,000. Its original
publication reported a minimum-over-samples, graph-aligned
interatomic-distance metric and coverage, so our single-sample RMSD is
not a reproduction of that published value. Supplementary Table S1 lists
the local and original-study protocols separately.

Paired comparisons used two-sided Wilcoxon signed-rank tests on
per-reaction errors, and monotonic associations used Spearman rank
correlation. All thresholds were fixed before held-out analysis.
Reactions that failed to reach a target remained in every denominator.
No held-out reaction contributed to checkpoint selection, threshold
setting or preprocessing decisions.

\subsection{Quantum-chemical saddle-point and energy-profile
validation}\label{quantum-chemical-saddle-point-and-energy-profile-validation}

Each held-out MARC-Loc prediction initialized a fresh Gaussian 16
transition-state optimization at $\omega$B97X/6-31G(d). A predefined primary
protocol was followed by one predefined rescue protocol for
non-converged cases. Rescue calculations restarted from the original
MARC-Loc geometry, not the final geometry of the failed run. The primary
protocol converged for 367 reactions and rescue recovered 40, giving 407
converged optimizations. Supplementary Methods provide the complete
route sections and charge, multiplicity, convergence and restart
settings.

Every converged structure underwent a frequency calculation at the same
level. A significant imaginary mode was defined as a frequency below -20
cm\textsuperscript{-1}; exactly one such mode classified a structure as
a frequency-confirmed first-order saddle-point candidate. After mass
weighting and removal of rigid translations and rotations, the unstable
direction was compared sign-invariantly with the local reference IRC
tangent. A separate score compared mode-induced distance changes for
atom pairs that differ between the endpoints. These tests assess local
saddle character and unstable-direction consistency. They do not
establish connection to the intended reactant and product basins, which
would require bidirectional IRC calculations.

For energy-profile analysis, MARC-Field was queried at every stored
coordinate of each held-out reference IRC. Gaussian single-point
energies at $\omega$B97X/6-31G(d) were then calculated for all generated
structures; all 53,888 calculations terminated normally. Each reference
and predicted profile was shifted independently by its own minimum:

\begin{equation*}
E_i^{\mathrm{rel}}=E_i-\min_{1\le j\le K}E_j.
\end{equation*}

Here \emph{E\textsubscript{i}} is the single-point energy of sampled
geometry \emph{i} and
$E_i^{\mathrm{rel}}$ its
value after subtracting the minimum within the same profile. \emph{K} is
the number of structures in the profile. Independent shifting removes
arbitrary absolute offsets, so the comparison measures relative profile
shape, span and peak position. We evaluated matched-coordinate profile
RMSE, energy span, endpoint-referenced forward and reverse barrier-like
descriptors and the coordinate of the highest-energy sampled structure.
These are descriptors of single-point profiles at fixed geometries, not
activation energies or force-converged minimum-energy paths.
Representative cases were selected only after aggregate analysis and did
not affect training, checkpoint selection or inference. Supplementary
Methods give the Gaussian settings.

\subsection{NEB initialization, budget accounting and
runtime}\label{neb-initialization-budget-accounting-and-runtime}

The NEB benchmark used 11-image bands for 100 predefined paired
reactions. For each MARC-Field band, the model was first queried at 11
uniformly spaced coordinates and a Gaussian single-point energy was
calculated at each structure. The highest-energy coordinate was refined
by a three-point quadratic fit when both neighbours were valid;
otherwise the discrete maximum was retained. The estimate was restricted
to {[}0.05, 0.95{]}. We then remapped a fixed 11-coordinate template so
that its central image coincided with this reaction-specific peak while
preserving \emph{u} = 0 and \emph{u} = 1. The control band used IDPP
interpolation implemented in the Atomic Simulation Environment (ASE)
{[}11,30{]} between the same endpoints. The remapping stretches or
compresses each half of the learned path without moving its endpoints.

NEB calculations used ASE {[}30{]} with Gaussian energies and forces at
$\omega$B97X/6-31G(d), the improved-tangent formulation {[}7{]}, spring
constant 0.1 and the FIRE optimizer. Setting the stopping-force
threshold to zero enforced exactly 100 optimizer steps for both arms.
The phase schedules differed: MARC-Field bands used 100 climbing-image
steps {[}8{]}, whereas IDPP bands used 30 ordinary NEB steps followed by
70 climbing-image steps.

The arms were not matched in total electronic-structure cost. The 11
single-point calculations used to centre each MARC-Field band were
initializer-construction costs outside the nominal 100-step budget, and
IDPP used no analogous scan. Thus, the matched budget covers only
optimizer steps after band construction. NEB-only timing excluded the
scan; an end-to-end parallel-time proxy added the longest of the 11
concurrently launched scan jobs. We did not reconstruct the exact number
of Gaussian calls to first target attainment, and we did not evaluate an
IDPP arm with the same scan, peak-centring and phase schedule. The
benchmark therefore compares the two predefined workflows, not the
isolated initializers at equal total cost.

A reaction was retained only when both arms passed checks for identity,
readable traces, 11-image integrity, finite energies and forces, fixed
endpoints and correct pairing. The benchmark used the first 100 valid
pairs in an order fixed before outcome analysis. The geometry target was
centre-image RMSD $\leq$ 0.10 \AA{} and the force target was maximum centre-image
atomic force $\leq$ 0.05 eV \AA{}\textsuperscript{-1}. Reactions that did not
reach a target remained in the denominator.

Neural inference was timed in 32-bit floating-point precision on one
NVIDIA RTX A6000 GPU after two warm-up passes. The complete
MARC-Field-to-MARC-Loc pipeline required a median of 122.9 ms per
reaction. Supplementary Fig. S11 reports component timing and scaling
with path resolution and molecular size.

\section{Data availability}\label{data-availability}

Transition1x is publicly available at
\url{https://gitlab.com/matschreiner/Transition1x}. T1x-IRC-8K, comprising the
8,209 curated reactions and 1,088,725 path-resolved geometries, together
with the predefined 7,409/390/410 split files, five-fold assignments,
processed evaluation tables, and the source data for all main and
supplementary figures, will be made available at
\url{https://huggingface.co/datasets/mathboylinlin/T1x-IRC-8K} upon
acceptance of the manuscript.

\section{Code availability}\label{code-availability}

Training, inference, evaluation and figure-generation code for MARC-TS,
together with the trained MARC-Field, MARC-Loc and DirectTS checkpoints
and the configurations used for the reported held-out results, will be
released at \url{https://github.com/mathboylinlin/MARC-TS} upon
acceptance of the manuscript.

\section{Acknowledgements}\label{acknowledgements}

This work was supported by the National Natural Science Foundation of
China (92470102, 52577152); the SEU Innovation Capability Enhancement Plan for Doctoral Students CXJH\_SEU
26186.

\section{Author contributions}\label{author-contributions}

Y.Y.: methodology, software, validation, data curation, investigation,
writing of original draft, review and editing, and visualization. L.Z.:
conceptualization, methodology, supervision, review and editing.

\section{Competing interests}\label{competing-interests}

The authors declare no competing interests.

\section*{References}
\begin{enumerate}[leftmargin=*,label={[\arabic*]},itemsep=0.25em,topsep=0.25em]
\small
\item Williams, I. H. Virtual transition states: making sense of multiple transition states in parallel and series. Chem. Soc. Rev. 54, 9145--9160 (2025).
\item Pinus, S., Genzling, J., Burai-Patrascu, M. \& Moitessier, N. Computational methods for asymmetric catalysis. Nat. Catal. 7, 1272--1287 (2024).
\item Joung, J. F., Casetti, N., Raghavan, P. \& Coley, C. W. An overview of reaction outcome prediction with physics-based and data-driven methods. Chem. Soc. Rev. 55, 6768--6813 (2026).
\item Butera, V. Density functional theory methods applied to homogeneous and heterogeneous catalysis: a short review and a practical user guide. Phys. Chem. Chem. Phys. 26, 7950--7970 (2024).
\item Koda, S. \& Saito, S. Locating transition states by variational reaction path optimization with an energy-derivative-free objective function. J. Chem. Theory Comput. 20, 2798--2811 (2024).
\item Beaglehole, I. W., Pemberton, M. J., Farrar, E. H. E. \& Grayson, M. N. Machine learning transition state geometries and applications in reaction property prediction. WIREs Comput. Mol. Sci. 15, e70025 (2025).
\item Henkelman, G. \& Jónsson, H. Improved tangent estimate in the nudged elastic band method for finding minimum energy paths and saddle points. J. Chem. Phys. 113, 9978--9985 (2000).
\item Henkelman, G., Uberuaga, B. P. \& Jónsson, H. A climbing image nudged elastic band method for finding saddle points and minimum energy paths. J. Chem. Phys. 113, 9901--9904 (2000).
\item E, W., Ren, W. \& Vanden-Eijnden, E. String method for the study of rare events. Phys. Rev. B 66, 052301 (2002).
\item Peters, B., Heyden, A., Bell, A. T. \& Chakraborty, A. A growing string method for determining transition states. J. Chem. Phys. 120, 7877--7886 (2004).
\item Smidstrup, S., Pedersen, A., Stokbro, K. \& Jónsson, H. Improved initial guess for minimum energy path calculations. J. Chem. Phys. 140, 214106 (2014).
\item Grambow, C. A., Pattanaik, L. \& Green, W. H. Reactants, products, and transition states of elementary chemical reactions based on quantum chemistry. Sci. Data 7, 137 (2020).
\item Schreiner, M., Bhowmik, A., Vegge, T., Busk, J. \& Winther, O. Transition1x---a dataset for building generalizable reactive machine learning potentials. Sci. Data 9, 779 (2022).
\item Thomas, N. et al. Tensor field networks: rotation- and translation-equivariant neural networks for 3D point clouds. Preprint at arXiv:1802.08219 (2018).
\item Satorras, V. G., Hoogeboom, E. \& Welling, M. E(n) equivariant graph neural networks. In Proc. 38th International Conference on Machine Learning 9323--9332 (PMLR, 2021).
\item Batzner, S. et al. E(3)-equivariant graph neural networks for data-efficient and accurate interatomic potentials. Nat. Commun. 13, 2453 (2022).
\item Batatia, I. et al. MACE: higher order equivariant message passing neural networks for fast and accurate force fields. In Advances in Neural Information Processing Systems 35, 11423--11436 (2022).
\item Jackson, R., Zhang, W. \& Pearson, J. TSNet: predicting transition state structures with tensor field networks and transfer learning. Chem. Sci. 12, 10022--10040 (2021).
\item Choi, S. Prediction of transition state structures of gas-phase chemical reactions via machine learning. Nat. Commun. 14, 1168 (2023).
\item Duan, C., Du, Y., Jia, H. \& Kulik, H. J. Accurate transition state generation with an object-aware equivariant elementary reaction diffusion model. Nat. Comput. Sci. 3, 1045--1055 (2023).
\item Kim, S., Woo, J. \& Kim, W. Y. Diffusion-based generative AI for exploring transition states from 2D molecular graphs. Nat. Commun. 15, 341 (2024).
\item Duan, C. et al. Optimal transport for generating transition states in chemical reactions. Nat. Mach. Intell. 7, 615--626 (2025).
\item Galustian, L., Mark, K., Karwounopoulos, J., Kovar, M. P.-P. \& Heid, E. GoFlow: efficient transition state geometry prediction with flow matching and E(3)-equivariant neural networks. Digital Discovery 4, 3492--3501 (2025).
\item Luo, Y., Gu, X. \& Sun, J. Generative flow model on distance geometry for predicting transition states of chemical reactions. Nat. Commun. (2026). doi:10.1038/s41467-026-74101-0.
\item Tuo, P., Chen, J. \& Li, J. Flow matching for reaction pathway generation. Nat. Commun. (2026). doi:10.1038/s41467-026-75654-w.
\item Ren, K. et al. Reactive machine learning potential for accelerating transition state search in organic synthesis. Nat. Commun. 17, 6253 (2026).
\item Schütt, K. T., Unke, O. T. \& Gastegger, M. Equivariant message passing for the prediction of tensorial properties and molecular spectra. In Proc. 38th International Conference on Machine Learning 9377--9388 (PMLR, 2021).
\item Kabsch, W. A solution for the best rotation to relate two sets of vectors. Acta Crystallogr. A 32, 922--923 (1976).
\item Frisch, M. J. et al. Gaussian 16, Revision C.01 (Gaussian, Inc., 2016).
\item Larsen, A. H. et al. The atomic simulation environment---a Python library for working with atoms. J. Phys.: Condens. Matter 29, 273002 (2017).
\end{enumerate}

\clearpage
\section*{Supplementary Information}
\addcontentsline{toc}{section}{Supplementary Information}
\setcounter{section}{0}
\setcounter{subsection}{0}
\renewcommand{\thesection}{S\arabic{section}}
\renewcommand{\thesubsection}{S\arabic{section}.\arabic{subsection}}
\renewcommand{\theHsection}{S\arabic{section}}
\renewcommand{\theHsubsection}{S\arabic{section}.\arabic{subsection}}

\section{Supplementary Methods}

The main Methods describe the experimental design, training and
validation needed to interpret the principal claims. The sections below
provide the mathematical definitions, implementation details and audit
procedures required to reproduce those analyses.

\subsection{Abbreviations and
notation}

Abbreviations used throughout are as follows: mass-weighted arc-length
reaction-coordinate framework for transition states (MARC-TS), its field
model (MARC-Field) and localizer (MARC-Loc); transition state (TS);
intrinsic reaction coordinate (IRC); root-mean-square deviation and
error (RMSD and RMSE); interquartile range (IQR); 90th and 95th
percentiles (P90 and P95); mean absolute error and percentage error (MAE
and MAPE); distance mean absolute error (D-MAE); density functional
theory and single-point calculation (DFT and DFT-SP); nudged elastic
band (NEB); image-dependent pair potential (IDPP); and Atomic Simulation
Environment (ASE). SE(3) denotes rigid rotations and translations in
three dimensions, PaiNN denotes an equivariant scalar-vector
message-passing architecture, Kabsch alignment denotes the optimal rigid
rotation minimizing coordinate RMSD, and EMA denotes an exponential
moving average. FP32 and bfloat16 specify numerical precision; GPU, SCF
and FIRE denote graphics processing unit, self-consistent field and Fast
Inertial Relaxation Engine, respectively. React-OT, OA-ReactDiff,
TSDiff, GoFlow and LearnTS are the external transition-state baselines.
ORACLE denotes a reference-informed post hoc diagnostic that is
unavailable at deployment.

Unless stated otherwise, \emph{\textbf{X}}\textsubscript{R} and
\emph{\textbf{X}}\textsubscript{P} denote reactant and product
Cartesian-coordinate matrices, \emph{\textbf{X}}\textsubscript{TS} the
transition-state coordinates, \textbf{Z} atomic identities, \emph{E} the
endpoint-union graph edges, \emph{N} the number of atoms,
\emph{m\textsubscript{i}} the mass of atom \emph{i}, \emph{u} the
normalized mass-weighted reaction coordinate, \emph{$\theta$} trainable model
parameters and $\Vert$$\cdot$$\Vert$\textsubscript{2} the Euclidean norm. Bold or
vector-valued coordinate symbols contain three Cartesian components per
atom.

Figure terminology is used consistently. A path slot is one fixed query
coordinate: `All 11 slots' uses \emph{u} = \{0.0, 0.1, \ldots, 1.0\};
`Interior 9 slots' uses \emph{u} = \{0.1, 0.2, \ldots, 0.9\}; and
`Central 5 slots' uses \emph{u} = \{0.30, 0.40, 0.50, 0.60, 0.70\}. This
evaluation subset differs from the MARC-Loc input window at \emph{u} =
\{0.40, 0.45, 0.50, 0.55, 0.60\}. The TS-associated frame is one
reference-linked evaluation, whereas complete-path RMSD is the
per-reaction mean over all 11 slots. An upper-tail exceedance curve
gives the fraction of reactions above an error threshold; hard-tail
denotes high-error reactions, not a separate split. In energy figures,
the transition window is \emph{u} = 0.40--0.60. For NEB, centre image
means the middle member of the 11-image band, highest-energy image means
the image with the largest sampled energy, and joint target requires the
same reaction to meet both geometry and force thresholds. Native
resolution means one query per stored IRC frame. IRC-ending and
static-endpoint refer to sampled IRC termini and independently optimized
endpoints, respectively.

\subsection{Dataset records, endpoint assignment and exact path
queries}

T1x-IRC-8K extends Transition1x {[}1{]}. Bidirectional IRC calculations
were initiated from all 10,073 reference transition states with the
$\omega$B97X functional and 6-31G(d) basis set, charge 0 and multiplicity 1.
The Gaussian route IRC=(CalcFC,MaxPoints=300,Recalc=2,StepSize=20)
computes an initial force-constant matrix, limits each direction to 300
points, requests periodic force-constant recalculation and sets
Gaussian\textquotesingle s IRC step-size parameter to 20. A trajectory
was retained only when both directions terminated normally and their
termini matched the designated reactant and product basins. Of 8,939
normally completed trajectories, 730 failed this endpoint-consistency
test. The final dataset contains 8,209 reactions and 1,088,725
geometries.

Each reaction record stores atomic identities, ordered atom mapping,
Cartesian coordinates, the reference transition-state index, electronic
energies, raw IRC direction and endpoint identities. Frame-level forces
and gradients are retained where available, together with validity
masks. We enforced reactant-first orientation and required finite
coordinates, constant atom identities and count, complete endpoint
ordering and at least one frame on each side of the reference transition
state. The neural models used only identities and geometries. The
dataset contains neutral C, H, N and O molecules, with a median of 14
atoms and trajectories ranging from tens to several hundred frames.

The final sampled IRC termini, rather than separately optimized minima,
are the native model boundaries. This preserves one trajectory
containing both boundaries, every intermediate structure and the
reference transition state. Supplementary Fig. S1 compares these termini
with independently optimized Transition1x structures. All formal queries
use the requested continuous coordinate directly; no nearest-frame
lookup is used for MARC-Loc windows, endpoint adaptation or path export.

\subsection{Mass-weighted coordinate and endpoint-union graph
details}

For consecutive path geometries
\emph{\textbf{X}}\textsuperscript{(\emph{k}-1)} and
\emph{\textbf{X}}\textsuperscript{(\emph{k})}, the mass-weighted
displacement, cumulative length and normalized coordinate are

\begin{equation*}
\Delta \ell_k=\left[\sum_{i=1}^{N}m_i\left\|\mathbf X_i^{(k)}-\mathbf X_i^{(k-1)}\right\|_2^2\right]^{1/2},\qquad
\ell_k=\sum_{q=1}^{k}\Delta\ell_q,\qquad
u_k=\frac{\ell_k}{\ell_K}.
\end{equation*}

Here \emph{\textbf{X}\textsubscript{i}}\textsuperscript{(\emph{k})} is
the Cartesian position of atom \emph{i} in frame \emph{k},
\emph{m\textsubscript{i}} is its atomic mass and \emph{N} is the atom
count. $\Delta$\emph{l}\textsubscript{k} is the mass-weighted displacement
between consecutive frames, \emph{l\textsubscript{k}} the cumulative
length to frame \emph{k}, K the final frame and
\emph{u\textsubscript{k}} =
\emph{l\textsubscript{k}}/\emph{l}\textsubscript{K} the normalized
coordinate. Thus \emph{u}\textsubscript{0} = 0 and
\emph{u}\textsubscript{K} = 1, while \emph{u}\textsubscript{TS} is the
coordinate of the stored transition-state frame. Because the coordinate
accumulates mass-weighted geometric distance, it generally differs from
the fraction of the frame index.

The 8,209 reactions were partitioned deterministically into 7,409
training, 390 validation and 410 held-out reactions, with disjoint
reaction identifiers. Five fold-excluded MARC-Field models generated
out-of-fold paths for groups of 1,482, 1,482, 1,482, 1,482 and 1,481
training reactions. The final MARC-Field model trained on all 7,409
reactions generated validation paths for checkpoint selection and
held-out paths for final evaluation. The held-out set was accessed only
after checkpoints, preprocessing rules and thresholds had been fixed.

For each reaction, an undirected endpoint-union graph was constructed
from bonded or spatially local pairs in either endpoint. For an edge
(\emph{i}, \emph{j}), the relative vector and distance are

\begin{equation*}
\mathbf r_{ij}=\mathbf X_i-\mathbf X_j,\qquad d_{ij}=\|\mathbf r_{ij}\|_2.
\end{equation*}

For atoms \emph{i} and \emph{j}, \emph{\textbf{r}\textsubscript{ij}} is
the displacement from \emph{j} to \emph{i} and
\emph{d\textsubscript{ij}} =
$\Vert$\emph{\textbf{r}\textsubscript{ij}}$\Vert$\textsubscript{2} is its Euclidean
length. These vectors provide direction to the equivariant
message-passing layers. The endpoint-union graph contains a pair when it
is bonded or spatially local in either endpoint, allowing the same graph
to cover bonds that form or break.

Distances were expanded in 20 radial basis functions under a 10 \AA{}
cutoff. Every equivariant block maintained scalar features and
Cartesian-vector features. PaiNN-style scalar-vector message passing and
edge-conditioned equivariant attention were combined through learned
gates and residual updates. Coordinate-dependent messages use relative
vectors, ensuring translation invariance and proper-rotation
equivariance.

\subsection{Detailed MARC-Field implementation and production
training}

The shared endpoint encoder produces scalar-vector states

\begin{equation*}
(S_{\mathrm R},\mathbf V_{\mathrm R})=\operatorname{Enc}_{\theta}(\mathbf Z,\mathbf X_{\mathrm R},E),\qquad
(S_{\mathrm P},\mathbf V_{\mathrm P})=\operatorname{Enc}_{\theta}(\mathbf Z,\mathbf X_{\mathrm P},E).
\end{equation*}

Enc\emph{\textsubscript{$\theta$}} denotes the shared endpoint encoder with
parameters \emph{$\theta$}. \emph{S}\textsubscript{R} and
\emph{S}\textsubscript{P} are scalar latent states for reactant and
product, \emph{\textbf{V}}\textsubscript{R} and
\emph{\textbf{V}}\textsubscript{P} are the corresponding vector latent
states, \textbf{Z} contains atomic identities and \emph{E} is the
endpoint-union edge set. The same encoder parameters are used for both
endpoints.

At query coordinate u, the latent states are linearly mixed,

\begin{equation*}
S(u)=(1-u)S_{\mathrm R}+uS_{\mathrm P},\qquad
\mathbf V(u)=(1-u)\mathbf V_{\mathrm R}+u\mathbf V_{\mathrm P}.
\end{equation*}

\emph{S}(\emph{u}) and \emph{\textbf{V}}(\emph{u}) are the scalar and
vector latent states supplied to the decoder at query coordinate
\emph{u}. The weights 1-\emph{u} and \emph{u} recover the reactant state
at \emph{u}=0, the product state at \emph{u}=1 and a latent mixture at
intermediate coordinates; nonlinear decoding then permits a curved
Cartesian path.

The decoder begins from Cartesian endpoint interpolation and applies
three geometry-recycling updates,

\begin{equation*}
\mathbf X^{(0)}(u)=(1-u)\mathbf X_{\mathrm R}+u\mathbf X_{\mathrm P},\qquad
\mathbf X^{(q)}(u)=\mathbf X^{(q-1)}(u)+\Delta\mathbf X^{(q)},\quad q=1,2,3.
\end{equation*}

\emph{\textbf{X}}\textsuperscript{(0)}(\emph{u}) is the Cartesian
endpoint interpolation used only to initialize decoding. At recycling
steps \emph{q} = 1, 2 and 3, the decoder predicts a correction
$\Delta$\emph{\textbf{X}}\textsuperscript{(\emph{q})} and updates the geometry
to \emph{\textbf{X}}\textsuperscript{(\emph{q})}(\emph{u}). Edge
distances and directions are recomputed after each update, so later
corrections depend on the current structure.

The production MARC-Field model uses latent dimension 512, eight encoder
blocks, eight decoder blocks, 16 attention heads, 20 radial basis
functions, a 10 \AA{} cutoff, a maximum drop-path rate of 0.20 and three
recycling updates. The query coordinate controls continuous latent
mixing, while nonlinear, geometry-dependent decoding permits a curved
Cartesian path.

On each training visit, one native IRC frame was sampled uniformly from
a reaction. Its exact coordinate \emph{u\textsubscript{i}} was the query
and its geometry \emph{\textbf{X}\textsubscript{i}} the target. Smooth
L1 coordinate losses with \emph{$\beta$} = 0.1 were applied to all three
recycling outputs with weights 0.5, 0.5 and 1.0. Endpoint structures
supplied boundary constraints and interior structures supplied path
supervision. No energies, forces or gradients were used.

Cross-fitting folds were generated with seed 42. Each fold model and the
final full-training model used at most 600 epochs, batch size 32, AdamW,
maximum learning rate 4$\times$10\textsuperscript{-4}, weight decay 0.05, a
OneCycle schedule with a 30\% increasing phase, gradient clipping at
0.5, EMA decay 0.999 and FP32 arithmetic. Validation occurred after
epoch 1 and every five epochs. Reference frames nearest \emph{u} = 0.1,
0.3, 0.5, 0.7 and 0.9 were queried at their actual coordinates, and the
EMA checkpoint with the lowest mean-squared coordinate error across
these positions was retained. Transition-state error was monitored but
not used for selection. Five training seeds were evaluated under the
same split and hyperparameters; seed 42 supplied the reported figures
and representative analyses.

Every cached training trajectory came from the fold checkpoint that
excluded that reaction. The final 7,409-reaction checkpoint generated
validation and held-out paths. Eleven-slot analyses used \emph{u} =
\{0.0, 0.1, \ldots, 1.0\}; native-resolution energy and timing analyses
used each stored coordinate array. Cartesian interpolation used the same
aligned endpoints and requested coordinates as MARC-Field.

\subsection{Detailed MARC-Loc features, objective and endpoint
adaptation}

The fixed local window contains queries at
\emph{u}=\{0.40,0.45,0.50,0.55,0.60\}. Let
\emph{\textbf{X}}\textsubscript{a} be the central anchor,
\emph{\textbf{X}}\textsubscript{l} and
\emph{\textbf{X}}\textsubscript{r} the left and right summaries and
\emph{\textbf{X}}\textsubscript{m} the window mean. The eight
equivariant vector channels are

\begin{align*}
\mathbf d_1&=\mathbf X_a-\mathbf X_{\mathrm R}, &
\mathbf d_2&=\mathbf X_{\mathrm P}-\mathbf X_a, &
\mathbf d_3&=\mathbf X_{\mathrm P}-\mathbf X_{\mathrm R}, &
\mathbf d_4&=\mathbf X_m-\mathbf X_a,\\
\mathbf d_5&=\mathbf X_l-\mathbf X_a, &
\mathbf d_6&=\mathbf X_r-\mathbf X_a, &
\mathbf d_7&=\mathbf X_r-\mathbf X_l, &
\mathbf d_8&=\mathbf X_l+\mathbf X_r-2\mathbf X_a.
\end{align*}

The eight vector channels separate endpoint and local-window
information: \emph{\textbf{d}}\textsubscript{1} and
\emph{\textbf{d}}\textsubscript{2} connect the anchor to the reactant
and product; \emph{\textbf{d}}\textsubscript{3} spans the endpoints;
\emph{d}\textsubscript{4} compares the window mean with the anchor;
\emph{\textbf{d}}\textsubscript{5} and
\emph{\textbf{d}}\textsubscript{6} compare the left and right summaries
with the anchor; \emph{\textbf{d}}\textsubscript{7} =
\emph{\textbf{X}}\textsubscript{r} - \emph{\textbf{X}}\textsubscript{l}
is tangent-like; and \emph{d}\textsubscript{8} =
\emph{\textbf{X}}\textsubscript{l} + \emph{\textbf{X}}\textsubscript{r}
- 2\emph{\textbf{X}}\textsubscript{a} is curvature-like. Twenty-four
scalar descriptors encode coordinates, window span, vector norms and
directional relationships. Atom embeddings and scalar descriptors
initialize scalar features, while learned coefficients combine the eight
channels into vector features. Eight SE(3)-equivariant blocks {[}6{]}
operate on the endpoint-union graph using anchor-based edge directions.

The final prediction is the bounded residual

\begin{equation*}
\mathbf X_{\mathrm{TS}}=\mathbf X_a+\Delta\mathbf X,\qquad
\Delta\mathbf X=\delta_{\max}\tanh\!\left(\frac{\Delta\mathbf X_{\mathrm{raw}}}{\delta_{\max}}\right),
\qquad \delta_{\max}=0.80\,\text{\AA}.
\end{equation*}

\emph{\textbf{X}}\textsubscript{a} is the central MARC-Field anchor,
$\Delta$\emph{\textbf{X}}\textsubscript{raw} is the unconstrained localizer
output and tanh acts elementwise. The bound \emph{$\delta$}\textsubscript{max}
= 0.80 \AA{} limits each Cartesian correction before it is added to
\emph{\textbf{X}}\textsubscript{a}, making MARC-Loc a residual refiner
rather than an independent geometry generator.

The production architecture uses latent dimension 384, eight residual
blocks, twelve attention heads, twenty radial basis functions, a 10 \AA{}
cutoff and maximum drop-path regularization rate 0.20. The
molecule-balanced objective is

\begin{equation*}
L_{\mathrm{Loc}}
=L_{\mathrm{atom}}+0.20L_{\mathrm{aux}}+0.01L_{\mathrm{pair}}
+3.0L_{\mathrm{RC\text{-}atom}}+0.5L_{\mathrm{RC\text{-}pair}}.
\end{equation*}

\emph{L}\textsubscript{atom} is the molecule-balanced all-atom
coordinate loss, \emph{L}\textsubscript{aux} an auxiliary atom-subset
loss, \emph{L}\textsubscript{pair} the all-pair distance loss,
\emph{L}\textsubscript{RC-atom} the reaction-centre atom loss and
\emph{L}\textsubscript{RC-pair} the endpoint-changing-pair loss. The
coefficients 0.20, 0.01, 3.0 and 0.5 are fixed training weights.
Reaction-centre terms receive the greatest emphasis because they
describe atoms and pairs directly involved in bond formation or
cleavage.

Atom-wise and pair-distance terms use Smooth L1 loss with \emph{$\beta$} =
0.05 and are averaged within each molecule before cohort averaging. The
auxiliary subset is defined by internal type index \textgreater{} 1; in
the production C/H/O/N encoding it contains H, O and N and is therefore
not a heavy-atom subset. All atom pairs are retained because the
implementation cap of 96 atoms exceeds every reaction size. The
reaction-centre atom mask contains changing-bond atoms and one
endpoint-neighbour shell; the pair term uses the changing pairs.
Residual-displacement regularization and a separate reaction-centre
auxiliary-atom term have zero weight in the production model.

Training supplied one clean, fixed, cross-fitted window per reaction per
epoch, with no coordinate noise or window jitter. MARC-Loc used at most
100 epochs, training/evaluation batches of 8/32, AdamW, maximum and
lower reference learning rates of 6$\times$10\textsuperscript{-5} and
5$\times$10\textsuperscript{-7}, a OneCycle schedule with a 15\% increasing
phase, weight decay 0.005, gradient clipping at 3.0, EMA decay 0.999,
maximum drop-path rate 0.20 and bfloat16 autocasting. Validation
occurred after epoch 1 and every two epochs; the selected EMA checkpoint
minimized mean all-atom proper-Kabsch RMSD on the validation set.
Training-loss early stopping was not used.

Endpoint-domain adaptation replaced sampled IRC termini with
independently optimized Transition1x structures while retaining the
predefined split and cross-fitted provenance. Exact five-frame windows
were queried without nearest-frame substitution. MARC-Loc was
initialized from the frozen native-endpoint EMA checkpoint; the
optimizer, scheduler and epoch counter were restarted, and fine-tuning
used only adapted windows. Validation selected epoch 4. Held-out mean,
median and P90 RMSDs were 0.193377, 0.135501 and 0.454071 \AA{}. This result
measures fine-tuned domain adaptation, not zero-shot transfer.

\subsection{Geometry metrics, statistics and baseline
implementation}

For predicted coordinates
$\hat{\mathbf X}$ and
reference coordinates \emph{\textbf{X}}, structures are centred and
aligned by the optimal proper Kabsch rotation
\emph{R}\textsuperscript{$\star$} with
det(\emph{R}\textsuperscript{$\star$})=+1{[}2{]}. The all-atom metric is

\begin{equation*}
\operatorname{RMSD}(\hat{\mathbf X},\mathbf X)=
\left[\frac{1}{N}\sum_{i=1}^{N}
\left\|R^{\star}(\hat{\mathbf X}_i-\bar{\hat{\mathbf X}})-(\mathbf X_i-\bar{\mathbf X})\right\|_2^2
\right]^{1/2}.
\end{equation*}

In the RMSD equation, \textbf{X}\emph{\textsubscript{i}} and
$\hat{\mathbf X}_i$ are the
reference and predicted coordinates for atom \emph{i},
$\bar{\mathbf X}$ and
$\bar{\hat{\mathbf X}}$ are
their centroids, \emph{R}\textsuperscript{$\star$} is the optimal proper
rotation and \emph{N} is the atom count. The determinant constraint
det(\emph{R}\textsuperscript{$\star$})=+1 excludes mirror reflection. The
denominator is \emph{N} atoms, not 3\emph{N} Cartesian components, so
the metric is an atom-wise geometric RMSD after rigid-body alignment.
Same-element rematching precedes alignment where generation can permute
equivalent atoms. MARC-TS, DirectTS and LearnTS{[}8{]} use proper
rotations without reflection; React-OT{[}7{]} and GoFlow{[}10{]} permit
reflection under their frozen local protocols, and OA-ReactDiff{[}9{]}
uses method-specific rematching, reflection and a 1 \AA{} per-reaction cap
before aggregation. Path RMSDs are summarized over all positions,
interior positions, the central five-point region and the
reference-TS-associated frame.

DirectTS uses the endpoint-union graph and MARC-Field endpoint backbone
as an isolated transition-state predictor. It receives only reactant and
product coordinates and atomic identities, with no intermediate
structures or five-frame window. The control uses the same split, latent
dimension 512, eight encoder and decoder blocks, 16 attention heads and
three recycling updates. Training used at most 600 epochs, batch size
32, AdamW, maximum learning rate 4$\times$10\textsuperscript{-4}, weight decay
0.05, a OneCycle schedule, gradient clipping at 0.5, EMA decay 0.999 and
FP32 arithmetic. The selected checkpoint minimized validation EMA mean
proper-Kabsch transition-state RMSD.

React-OT {[}7{]}, LearnTS {[}8{]}, OA-ReactDiff {[}9{]}, GoFlow {[}10{]}
and TSDiff {[}11{]} were evaluated from frozen local checkpoints.
LearnTS used epoch 1453. TSDiff used iteration 67,000, one sample per
reaction and 5,000 Langevin steps; same-element rematching followed by
proper-Kabsch alignment without reflection gave mean, median and P90
RMSDs of 0.563345, 0.604037 and 0.927759 \AA{}. A separate audit using exact
automorphisms of the labelled condensed reaction graph gave a
paper-style distance error of 0.140519 \AA{}, with 48.29\% and 71.95\%
coverage below 0.1 and 0.2 \AA{}. Because the public implementation does not
include the authors\textquotesingle{} atom-index-alignment evaluator,
this is a paper-faithful reconstruction rather than a byte-for-byte
reproduction. Supplementary Table S1 records all method-specific
conventions.

\subsection{Quantum-chemical validation and energy-profile
implementation}

Every held-out MARC-Loc prediction was submitted as a fresh Gaussian
16{[}3{]} calculation at $\omega$B97X/6-31G(d), charge 0 and multiplicity 1.
The primary route was

Opt=(TS,CalcFC,NoEigenTest,Cartesian,MaxStep=5,MaxCycles=300)

SCF=(XQC,MaxCycle=512) Int=UltraFine NoSymm

In this Gaussian route, TS requests transition-state optimization;
CalcFC computes an initial force-constant matrix; NoEigenTest disables
the initial Hessian eigenvalue test; Cartesian selects Cartesian
optimization coordinates; and MaxStep and MaxCycles set the step and
cycle limits. XQC enables a quadratic-convergence fallback for the
self-consistent-field procedure, MaxCycle=512 sets its cycle limit,
Int=UltraFine selects the integration grid and NoSymm disables symmetry
constraints.

Normal termination after transition-state optimization defined
convergence. The primary route converged for 367 of 410 reactions. The
43 failures restarted from the original MARC-Loc geometry, not a failed
intermediate, using Opt=(TS,CalcFC,NoEigenTest,MaxCycles=200) with the
same electronic-structure, integration and SCF settings. Forty
additional reactions converged, giving 407 of 410.

Frequency calculations used Freq Geom=AllCheck Guess=Read, reusing the
optimized geometry and wavefunction. A mode below -20
cm\textsuperscript{-1} was considered significantly imaginary, and
exactly one such mode defined a frequency-confirmed first-order
saddle-point candidate. The unstable mode was mass weighted, projected
to remove rigid translations and rotations and compared sign-invariantly
with the local reference IRC tangent. A second score compared signs of
mode-induced distance changes for endpoint-changing pairs. The cutoff
excludes near-zero numerical modes. These tests assess local saddle
character and direction, not connectivity to the intended endpoints.

For energy profiles, MARC-Field was queried at every native coordinate
of the 410 held-out reactions. Gaussian single-point energies were
evaluated at $\omega$B97X/6-31G(d) with SCF=(XQC,MaxCycle=512) and NoSymm; all
53,888 calculations terminated normally. Energies were converted from
hartree to kcal mol\textsuperscript{-1}, and each profile was shifted by
its own minimum:

\begin{equation*}
E_i^{\mathrm{rel}}=E_i-\min_{1\le j\le K}E_j.
\end{equation*}

\emph{E\textsubscript{i}} is the electronic energy of sampled geometry
\emph{i}, \emph{K} is the number of structures in the profile and
$E_i^{\mathrm{rel}}$ is the
energy after subtracting the minimum of that profile. Reference and
predicted profiles are normalized independently, so the comparison
measures relative shape, span and peak position rather than an arbitrary
absolute offset.

Metrics comprised matched-coordinate profile RMSE, energy span, forward
and reverse endpoint-referenced barrier-like descriptors and
highest-energy coordinate. These single-point quantities are not
force-converged NEB paths, activation energies or frequency-confirmed
transition states. Representative reactions were selected only after
aggregate analysis. Reaction 571 contains a carbene-like C2 product
centre with non-bonding electrons; its elevated product-side energy is
attributed to unusual electronic structure rather than a generic broad
transition region.

\subsection{NEB construction, budget accounting and cohort
audit}

For each reaction, MARC-Field was queried at \emph{u} = \{0.0, 0.1,
\ldots, 1.0\} and 11 Gaussian single-point energies were evaluated. The
highest-energy coordinate was selected and, when both neighbouring
energies were valid, refined by a three-point quadratic fit. Otherwise,
the discrete maximum was retained. The resulting
\emph{u}\textsubscript{peak} was restricted to {[}0.05, 0.95{]}.

The fixed abstract template
\emph{t}=\{0.00,0.10,0.25,0.37,0.45,0.50,0.55,0.63,0.75,0.90,1.00\} was
mapped by

\begin{align*}
u(t)&=2u_{\mathrm{peak}}t, && t\le 0.5,\\
u(t)&=u_{\mathrm{peak}}+2(1-u_{\mathrm{peak}})(t-0.5), && t>0.5.
\end{align*}

The template coordinate \emph{t} indexes the fixed 11-image pattern and
\emph{u}(\emph{t}) is the mass-weighted coordinate queried from
MARC-Field. The piecewise map rescales the reactant and product halves
separately, preserving \emph{u}(0) = 0 and \emph{u}(1) = 1 while
enforcing \emph{u}(0.5) = \emph{u}\textsubscript{peak}.

The remapped coordinates place the central image at \emph{the estimated
peak} while preserving both endpoints. The control used ASE IDPP
interpolation {[}4,5{]} between the same endpoints. Gaussian energies
and forces used \#p wb97x/6-31g(d) Force SCF=(XQC,MaxCycle=512) NoSymm,
charge 0 and multiplicity 1. Both arms used 11 images, fixed endpoints,
the improved tangent, spring constant 0.1, FIRE and exactly 100
optimizer steps. The schedules differed: MARC-Field used 100
climbing-image steps, whereas IDPP used 30 ordinary followed by 70
climbing-image steps.

The 11-point DFT scan is part of MARC-Field initializer construction but
lies outside the nominal 100-step NEB budget. Thus, `same 100-step
budget' means the same number of optimizer steps after image
construction, not the same total number of electronic-structure
calculations. NEB-only wall time excludes the scan. The MARC-Field
end-to-end parallel proxy adds the longest of the 11 concurrently
launched scan jobs; IDPP has no scan. Exact Gaussian calls to first
target attainment were not reconstructed and are not claimed. No IDPP
arm used the same scan, centring and phase schedule.

The selected-job tables contained 137 candidates in a fixed order. A
pair was valid only when both arms passed checks for reaction identity,
readable traces, 11-image integrity, finite energies and forces, fixed
endpoints and correct pairing. The first 100 valid pairs in the
pre-specified order formed the benchmark; outcomes did not determine
inclusion order. Metrics include final centre-image RMSD, centre-image
maximum atomic force, maximum projected NEB force, highest-energy-image
RMSD and highest-image relative-energy error. Unreached reactions remain
in every denominator. Supplementary Fig. S10 uses these 100 pairs and
reports geometry, force and joint-target fractions of 87\% versus 60\%,
90\% versus 47\% and 66\% versus 12\%. The symmetric paired gain is
(\emph{e}\textsubscript{IDPP} -
\emph{e}\textsubscript{MARC})/(\emph{e}\textsubscript{IDPP} +
\emph{e}\textsubscript{MARC}), so positive values favour MARC-Field.

\subsection{Runtime measurement}

Timing used FP32 on one NVIDIA RTX A6000 after two warm-up passes and
covered all 410 held-out reactions. Measurements included one-coordinate
MARC-Field inference, batched five-frame inference, MARC-Loc with a
precomputed window, the complete field-to-localizer pipeline and a
native-resolution path. Median end-to-end neural latency was 122.9 ms
per reaction. Full-path latency scales with the number of requested
coordinates; the five-frame window remains close to one-coordinate cost
because its queries are batched. Supplementary Fig. S11 reports
component distributions and scaling with resolution and molecular size.

\section{Supplementary Tables}

\subsection{Supplementary Table S1: extended baseline
context}

Supplementary Table S1 Published context and common-cohort
evaluations. Original-study values are shown only for context and are
not directly comparable across datasets, inputs or metrics. `This work'
denotes local evaluations on the predefined T1x-IRC-8K held-out set. P90
and P95 are the 90th and 95th percentiles.

\begin{longtable}[]{@{}
  >{\raggedright\arraybackslash}p{(\columnwidth - 4\tabcolsep) * \real{0.2674}}
  >{\raggedright\arraybackslash}p{(\columnwidth - 4\tabcolsep) * \real{0.3721}}
  >{\raggedright\arraybackslash}p{(\columnwidth - 4\tabcolsep) * \real{0.3605}}@{}}
\toprule\noalign{}
\begin{minipage}[b]{\linewidth}\raggedright
\textbf{Method and input}
\end{minipage} & \begin{minipage}[b]{\linewidth}\raggedright
\textbf{Original-study context}
\end{minipage} & \begin{minipage}[b]{\linewidth}\raggedright
\textbf{Local protocol in this work and mean / median / P90 RMSD (\AA{})}
\end{minipage} \\
\midrule\noalign{}
\endhead
\bottomrule\noalign{}
\endlastfoot
MARC-TS 3D R/P + path & T1x-IRC-8K; continuous path prediction, TS
localization and NEB initialization. & 7,409 training; 390 validation;
final evaluation on the 410-reaction held-out set. 0.1269 / 0.0902 /
0.2776. \\
DirectTS 3D R/P & Internal endpoint-only control. & Same reaction IDs
and reporting metric; no path window. 0.1424 / 0.1153 / -\/-. \\
React-OT {[}7{]} 3D R/P & Transition1x; published mean/median
0.103/0.053 \AA{} after OA-ReactDiff initialization and OT fine-tuning;
lower-theory pretraining gives a separate literature result. & Local
1,000-epoch reproduction; reflection permitted in the frozen local
protocol. 0.1604 / 0.1256 / 0.3295. \\
LearnTS {[}8{]} 3D R/P + interpolation & Distance MAE/MAPE 10.70
pm/3.407\%; 93.8\% saddle-point convergence. & Updated epoch 1453;
proper rotation; P95 0.433495 \AA{}; maximum 0.872852 \AA{}. 0.170813 / 0.140973
/ 0.329449. \\
OA-ReactDiff {[}9{]} 3D R/P & Transition1x median TS RMSD 0.08 \AA{};
reported barrier error 2.6 kcal mol\textsuperscript{-1} with
confidence-guided refinement. & Epoch 2,000 EMA; same-element
rematching; reflection allowed; per-reaction cap 1 \AA{}. 0.1891 / 0.0965 /
0.4699. \\
GoFlow {[}10{]} 2D graph & RDB7 reaction benchmark; published RMSD 0.170
$\pm$ 0.002 \AA{} and D-MAE 0.104 $\pm$ 0.002 \AA{}. & Epoch 650; 25 Euler steps;
10-sample median-reference aggregation; same-element rematching;
reflection allowed. 0.2957 / 0.2104 / 0.6770. \\
TSDiff {[}11{]} 2D graph & Stochastic diffusion from 2D molecular
graphs; the original study evaluates minimum-over-samples graph-aligned
interatomic-distance error and coverage rather than the present
single-sample RMSD aggregation. & Iteration 67,000 selected by
validation loss; one sample per reaction; 5,000 Langevin steps;
same-element rematching followed by proper Kabsch without reflection.
0.563345 / 0.604037 / 0.927759. Paper-metric audit: mean graph-aligned
distance error 0.140519 \AA{}; coverage below 0.1 \AA{} is 48.29\% and coverage
below 0.2 \AA{} is 71.95\%. \\
\end{longtable}

\subsection{Supplementary Table S2: production architecture and training
summary}

Supplementary Table S2 Production architecture and training
summary. Configuration of the MARC-Field and MARC-Loc production models,
optimization and checkpoint selection.

\begin{longtable}[]{@{}
  >{\raggedright\arraybackslash}p{(\columnwidth - 4\tabcolsep) * \real{0.2022}}
  >{\raggedright\arraybackslash}p{(\columnwidth - 4\tabcolsep) * \real{0.2078}}
  >{\raggedright\arraybackslash}p{(\columnwidth - 4\tabcolsep) * \real{0.5900}}@{}}
\toprule\noalign{}
\begin{minipage}[b]{\linewidth}\raggedright
\textbf{Setting}
\end{minipage} & \begin{minipage}[b]{\linewidth}\raggedright
\textbf{MARC-Field}
\end{minipage} & \begin{minipage}[b]{\linewidth}\raggedright
\textbf{MARC-Loc}
\end{minipage} \\
\midrule\noalign{}
\endhead
\bottomrule\noalign{}
\endlastfoot
Primary role & Continuous endpoint-conditioned path field & Five-frame
residual TS localizer \\
Latent dimension & 512 & 384 \\
Equivariant blocks & 8 encoder + 8 decoder & 8 residual blocks \\
Attention heads & 16 & 12 \\
Radial basis / cutoff & 20 / 10 \AA{} & 20 / 10 \AA{} \\
Geometry recycling & 3 decoder updates & Not applicable \\
Window / query coordinates & Arbitrary \emph{u} in {[}0,1{]} & 0.40,
0.45, 0.50, 0.55, 0.60 \\
Residual bound & Decoder displacement updates & Component-wise 0.80 \AA{} \\
Maximum drop-path rate & 0.20 & 0.20 \\
Training set & 7,409 reactions; five-fold cross-fitting plus final
full-training model & 7,409 reactions with cross-fitted MARC-Field
windows \\
Validation / held-out set & 390 / 410 reactions & 390 / 410 reactions \\
Training sample & One randomly sampled native IRC frame per reaction
visit & One clean fixed five-frame window per reaction per epoch \\
Coordinate noise / window jitter & Not used in frozen production
description & 0 / 0 \\
Loss summary & Three-recycle coordinate Smooth L1 loss (\emph{$\beta$} = 0.1;
weights 0.5, 0.5, 1.0) &
$L_{\mathrm{atom}}+0.20L_{\mathrm{aux}}+0.01L_{\mathrm{pair}}+3.0L_{\mathrm{RC\text{-}atom}}+0.5L_{\mathrm{RC\text{-}pair}}$ \\
Epoch budget & At most 600 & At most 100 \\
Train / evaluation batch & 32 / not separately batched for validation
export & 8 / 32 \\
Optimizer & AdamW & AdamW \\
Maximum / lower learning rate & 4 $\times$ 10\textsuperscript{-4} / OneCycle
minimum determined by schedule & 6 $\times$ 10\textsuperscript{-5} / 5 $\times$
10\textsuperscript{-7} \\
Weight decay & 0.05 & 0.005 \\
OneCycle increasing phase & 30\% & 15\% \\
Gradient-norm clipping & 0.5 & 3.0 \\
EMA decay & 0.999 & 0.999 \\
Numerical precision & FP32 & bfloat16 autocasting \\
Validation frequency & Epoch 1, then every 5 epochs & Epoch 1, then
every 2 epochs \\
Checkpoint selection & EMA five-position mean-squared coordinate error
on validation paths & EMA mean all-atom proper-Kabsch TS RMSD \\
Early stopping & None; fixed epoch budget & None; fixed epoch budget \\
Training provenance & Fold-excluded paths for all training reactions &
Fold-excluded MARC-Field windows \\
\end{longtable}

\section{Supplementary Figures}

\begin{figure}[H]
\centering
\includegraphics[width=\linewidth,height=0.62\textheight,keepaspectratio]{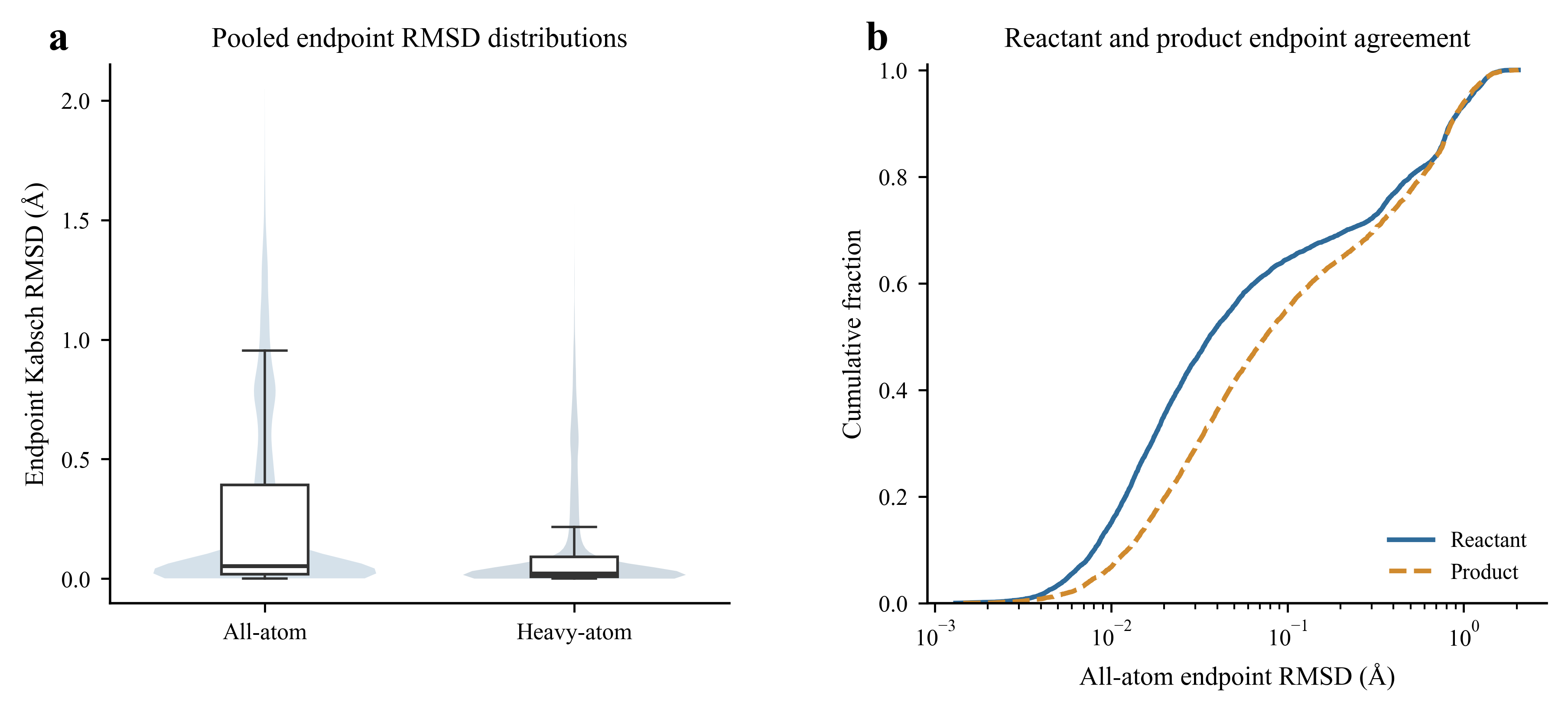}
\par\smallskip
\begin{minipage}{0.98\linewidth}\small
Supplementary Fig. S1 Geometric agreement between sampled IRC endpoints
and independently optimized endpoints. \textbf{a}, Pooled all-atom and
heavy-atom proper-Kabsch RMSD distributions. \textbf{b}, Cumulative
all-atom RMSD distributions for reactant and product endpoints. The
analysis contains 16,418 endpoints from 8,209 reactions and uses fixed
atom correspondence without endpoint swapping, atom permutation or
mirror reflection.
\end{minipage}
\end{figure}

\begin{figure}[H]
\centering
\includegraphics[width=\linewidth,height=0.62\textheight,keepaspectratio]{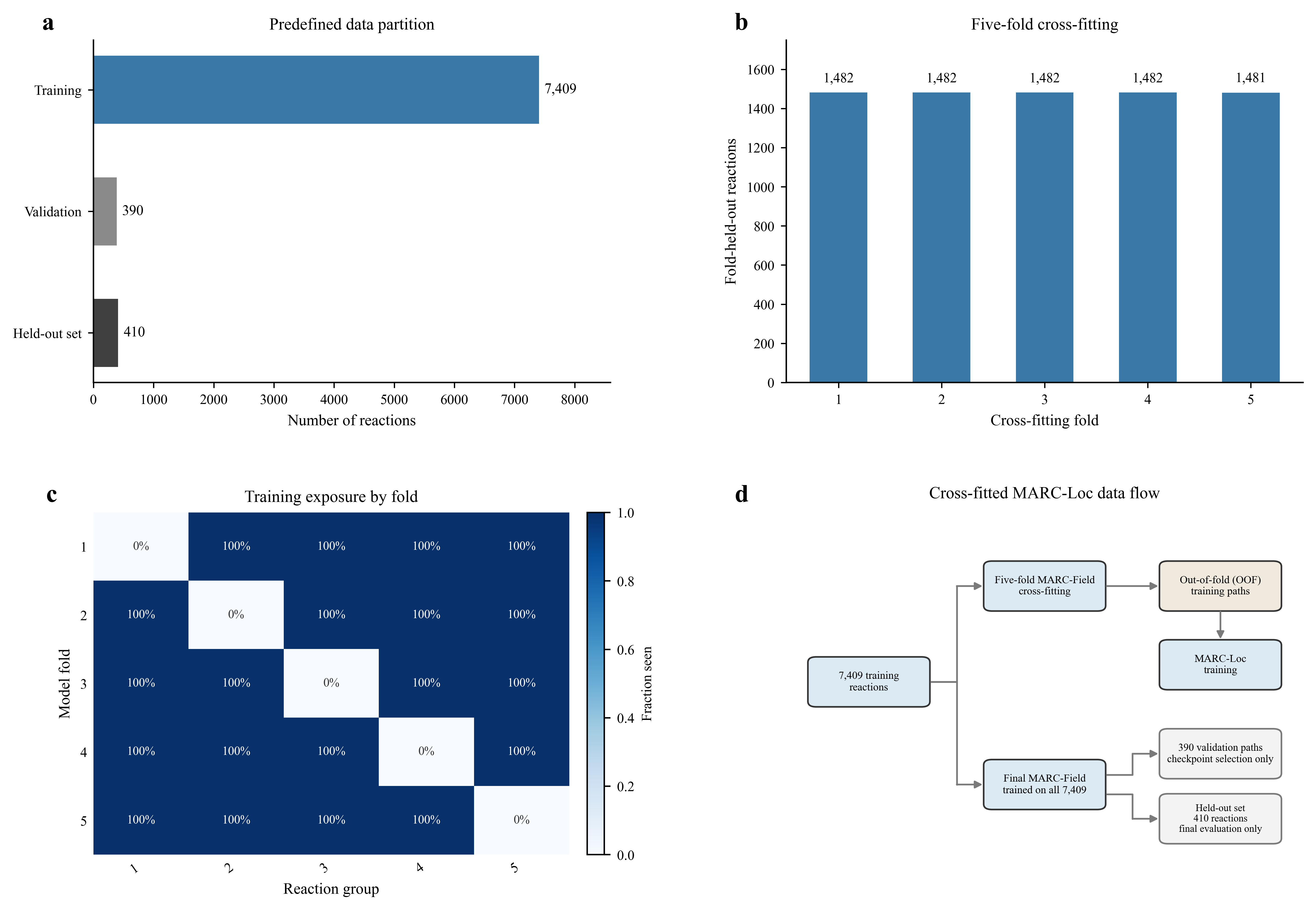}
\par\smallskip
\begin{minipage}{0.98\linewidth}\small
Supplementary Fig. S2 Predefined split and cross-fitted MARC-Loc data
flow. \textbf{a}, Partition into 7,409 training, 390 validation and 410
held-out reactions. \textbf{b}, Sizes of the five excluded groups used
for MARC-Field cross-fitting. \textbf{c}, Fold-exclusion matrix;
diagonal 0\% entries mark the group omitted from each fold model.
\textbf{d}, Cross-fitted MARC-Field models generate out-of-fold MARC-Loc
training paths, whereas the final model trained on all 7,409 training
reactions generates validation paths for checkpoint selection and
held-out paths for final evaluation.
\end{minipage}
\end{figure}

\begin{figure}[H]
\centering
\includegraphics[width=\linewidth,height=0.62\textheight,keepaspectratio]{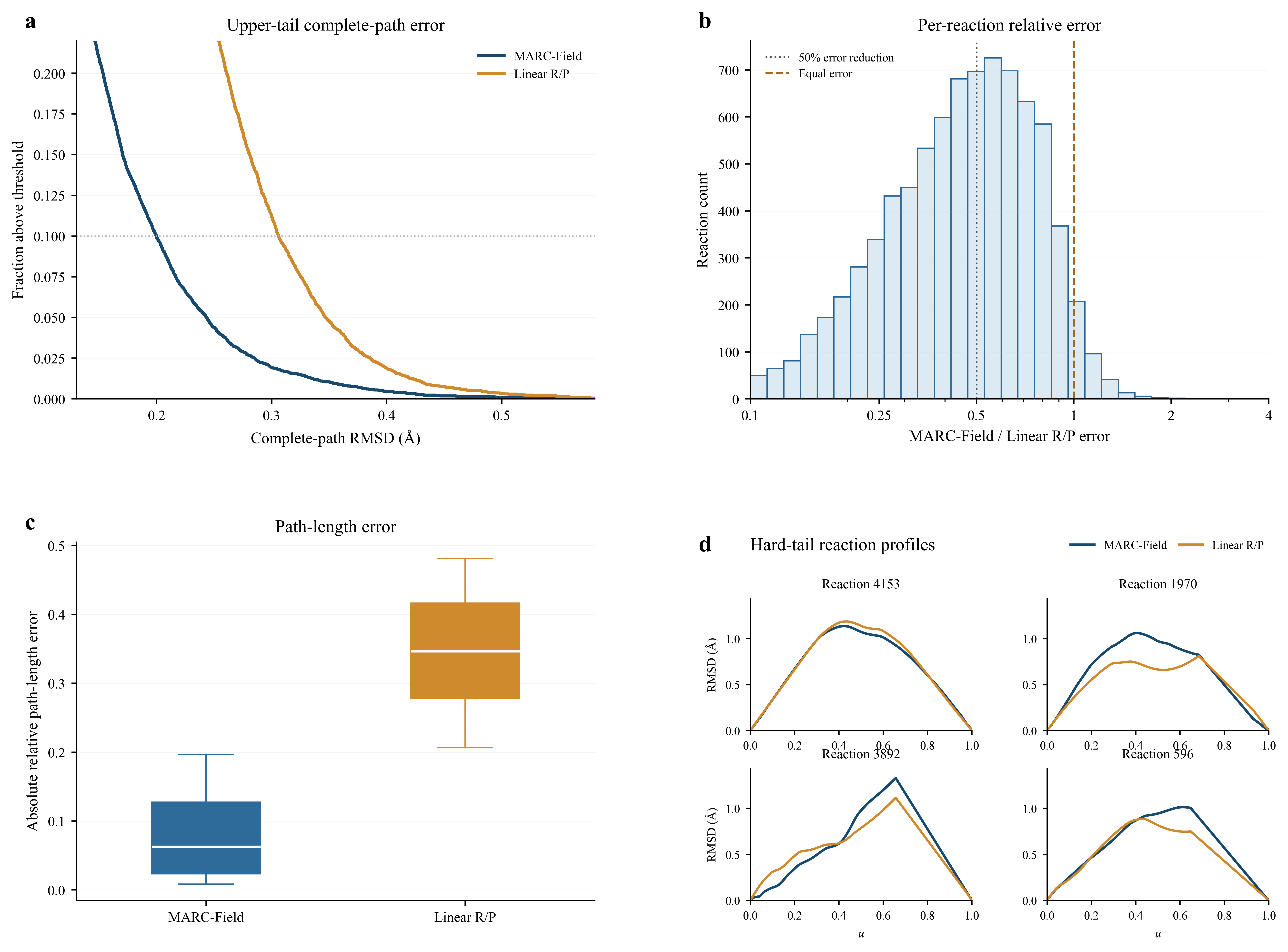}
\par\smallskip
\begin{minipage}{0.98\linewidth}\small
Supplementary Fig. S3 MARC-Field tail-error and path-length diagnostics.
\textbf{a}, Upper-tail exceedance curves of per-reaction complete-path
RMSD for MARC-Field and linear reactant/product interpolation.
\textbf{b}, Distribution of their complete-path error ratio; dotted and
dashed lines mark 50\% error reduction and equal error. \textbf{c},
Absolute relative mass-weighted path-length error. \textbf{d},
Coordinate-resolved errors for four reactions selected from the
MARC-Field high-error tail. Box plots show medians, interquartile ranges
and 10th--90th-percentile whiskers.
\end{minipage}
\end{figure}

\begin{figure}[H]
\centering
\includegraphics[width=\linewidth,height=0.62\textheight,keepaspectratio]{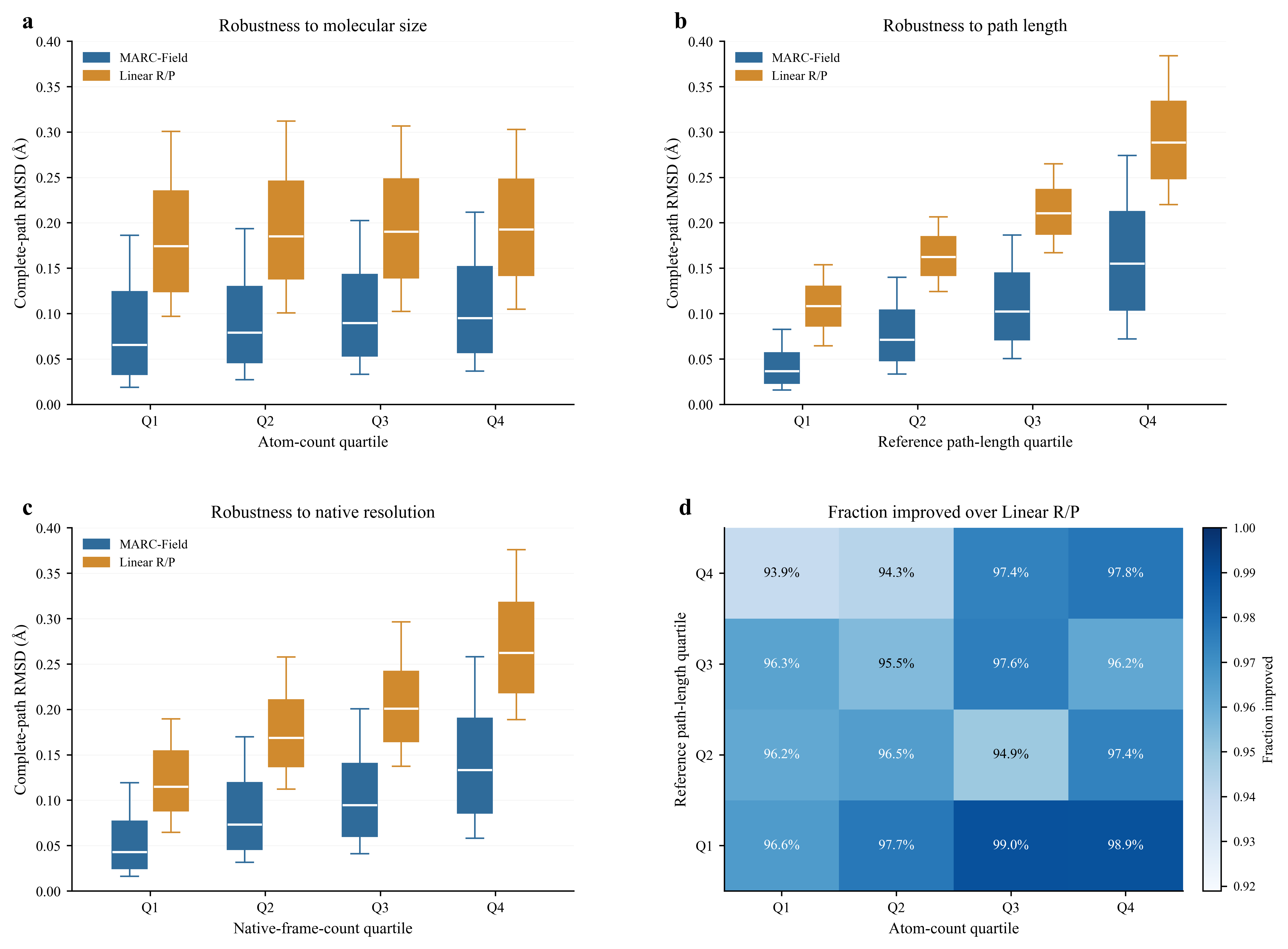}
\par\smallskip
\begin{minipage}{0.98\linewidth}\small
Supplementary Fig. S4 Path fidelity across reaction complexity.
\textbf{a--c}, Complete-path RMSD for MARC-Field and linear
reactant/product interpolation across quartiles of atom count, reference
path length and native IRC resolution. \textbf{d}, Fraction of reactions
for which MARC-Field has lower error across the joint atom-count and
path-length grid. Quartiles are defined from the corresponding
reaction-level distributions.
\end{minipage}
\end{figure}

\begin{figure}[H]
\centering
\includegraphics[width=\linewidth,height=0.62\textheight,keepaspectratio]{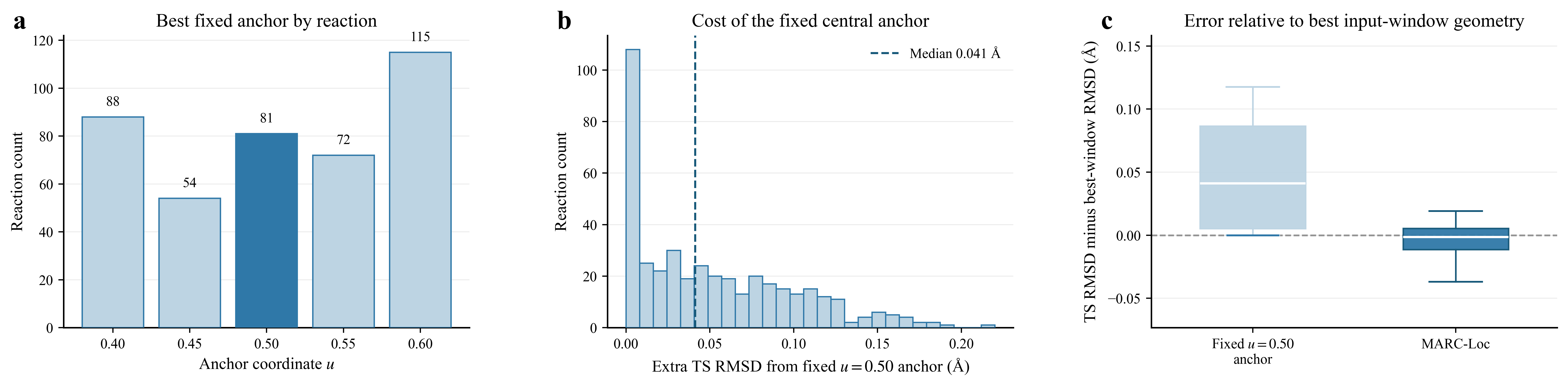}
\par\smallskip
\begin{minipage}{0.98\linewidth}\small
Supplementary Fig. S5 Reference-informed diagnostics of the MARC-Loc
input window. \textbf{a}, Coordinate that gives the lowest reference
transition-state RMSD for each reaction. \textbf{b}, Extra RMSD from
fixing the anchor at u = 0.50 rather than choosing the best of five
input coordinates; the dashed line marks the median. \textbf{c},
Transition-state RMSD relative to the best input geometry for the fixed
anchor and MARC-Loc; negative values indicate that MARC-Loc improves on
every input frame. These ORACLE quantities use the reference transition
state and are unavailable at deployment.
\end{minipage}
\end{figure}

\begin{figure}[H]
\centering
\includegraphics[width=\linewidth,height=0.62\textheight,keepaspectratio]{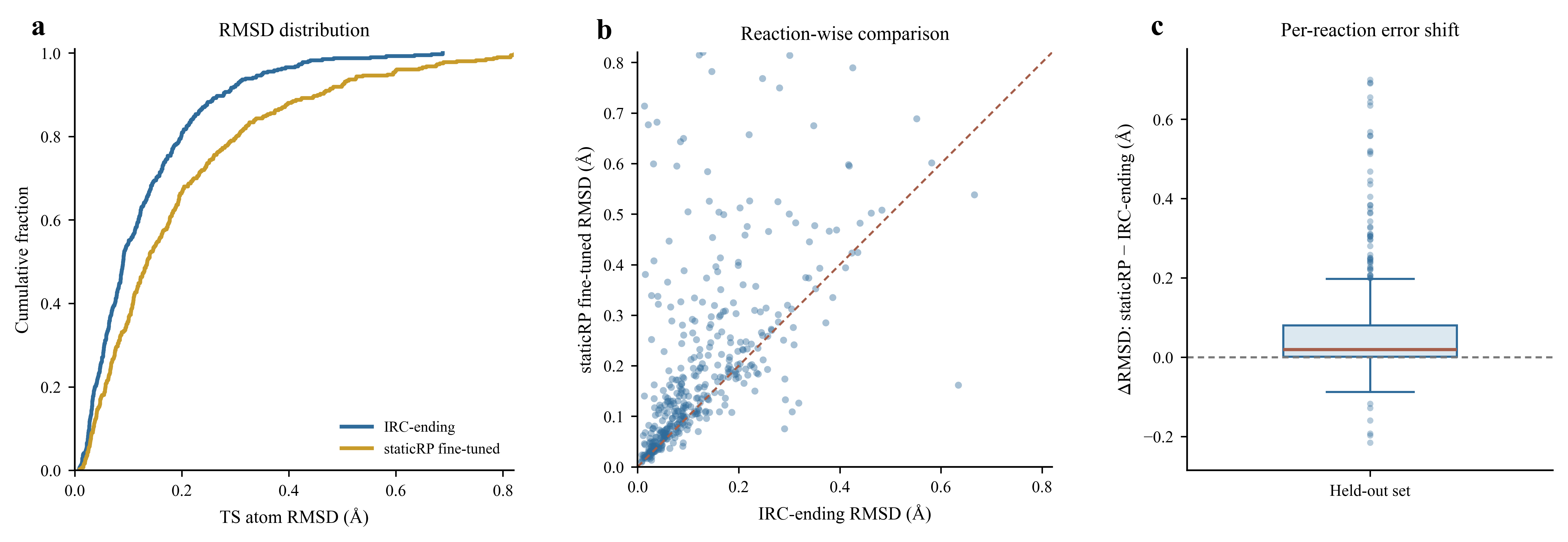}
\par\smallskip
\begin{minipage}{0.98\linewidth}\small
Supplementary Fig. S6 Fine-tuned adaptation to independently optimized
endpoints. \textbf{a}, Cumulative transition-state RMSD distributions
for sampled IRC endpoints and fine-tuned static endpoints. \textbf{b},
Reaction-wise comparison for the same 410 held-out reactions; the dashed
diagonal indicates equality. \textbf{c}, Distribution of the
per-reaction RMSD change, defined as static-endpoint minus IRC-endpoint
RMSD; the dashed line marks zero. The endpoint-adapted model was
fine-tuned and selected on validation data and is not a zero-shot
result.
\end{minipage}
\end{figure}

\begin{figure}[H]
\centering
\includegraphics[width=\linewidth,height=0.62\textheight,keepaspectratio]{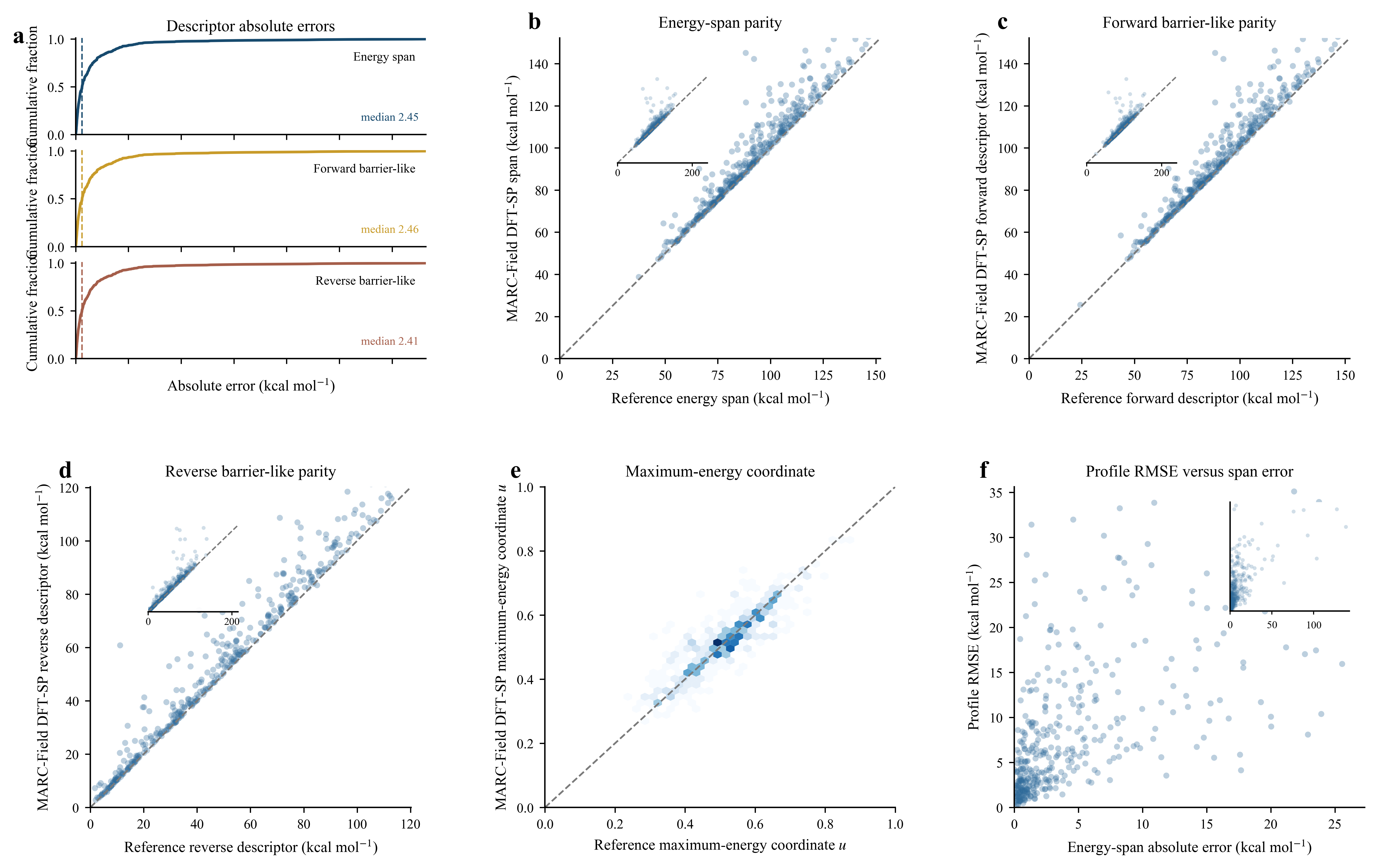}
\par\smallskip
\begin{minipage}{0.98\linewidth}\small
Supplementary Fig. S7 Fidelity of DFT single-point energy
descriptors. \textbf{a}, Cumulative absolute errors in energy span and
forward and reverse barrier-like descriptors; dashed lines mark medians.
\textbf{b--d}, Reference-versus-MARC-Field parity for the three
descriptors; insets retain full ranges. \textbf{e}, Reference versus
predicted maximum-energy coordinate. \textbf{f}, Relationship between
profile RMSE and energy-span error. All quantities come from
single-point profiles on fixed sampled geometries; the barrier-like
descriptors are not force-converged activation energies.
\end{minipage}
\end{figure}

\begin{figure}[H]
\centering
\includegraphics[width=\linewidth,height=0.62\textheight,keepaspectratio]{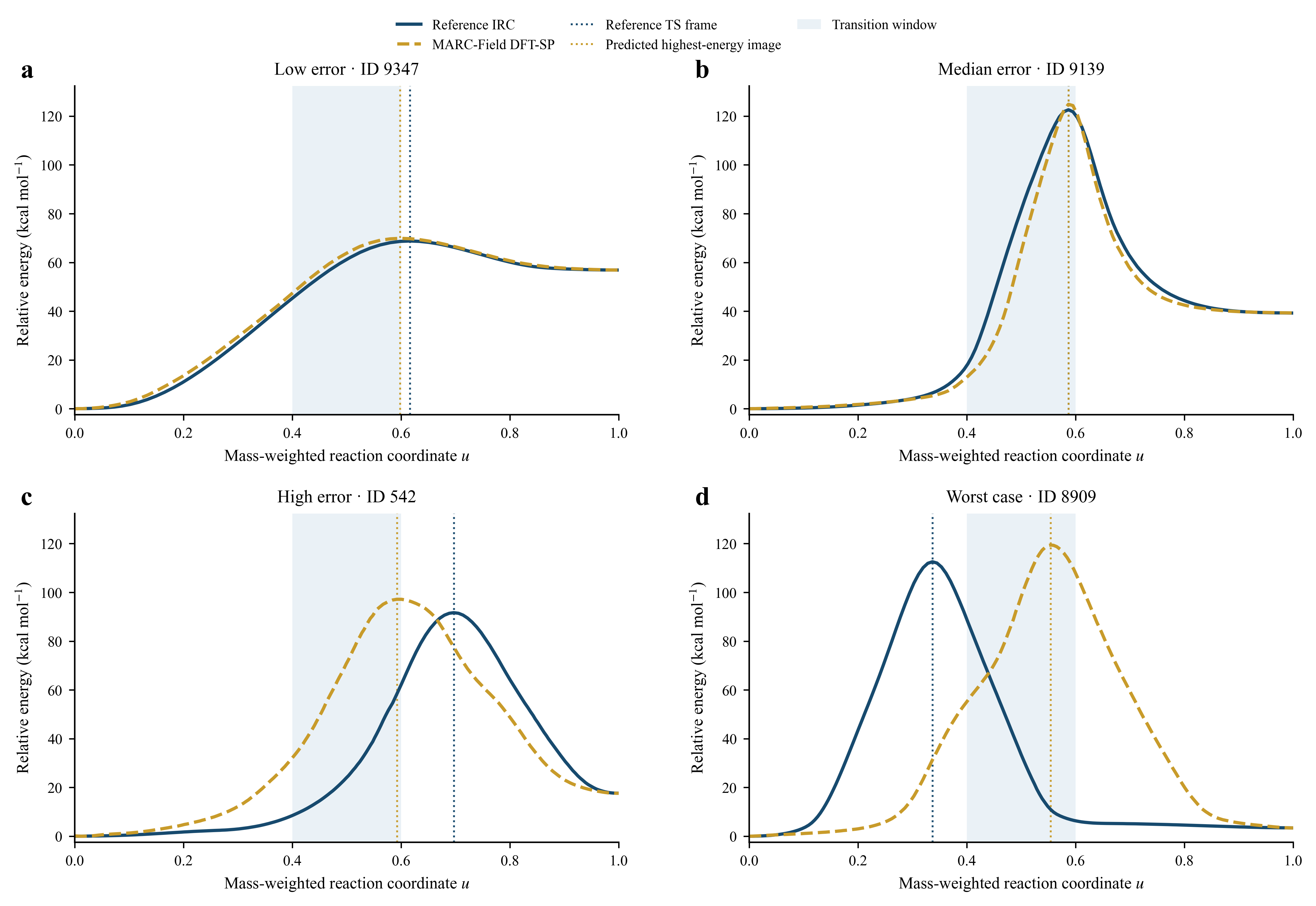}
\par\smallskip
\begin{minipage}{0.98\linewidth}\small
Supplementary Fig. S8 Representative DFT single-point energy profiles
across the error distribution. \textbf{a--d}, Reactions 9347, 9139, 542
and 8909 represent low, median, high and worst profile-RMSE regimes.
Solid blue and dashed ochre curves show reference IRC and MARC-Field
profiles. Vertical markers identify the reference transition-state frame
and predicted highest-energy image; shading marks \emph{u} = 0.40--0.60.
The examples were fixed from the error distribution after aggregate
analysis.
\end{minipage}
\end{figure}

\begin{figure}[H]
\centering
\includegraphics[width=\linewidth,height=0.62\textheight,keepaspectratio]{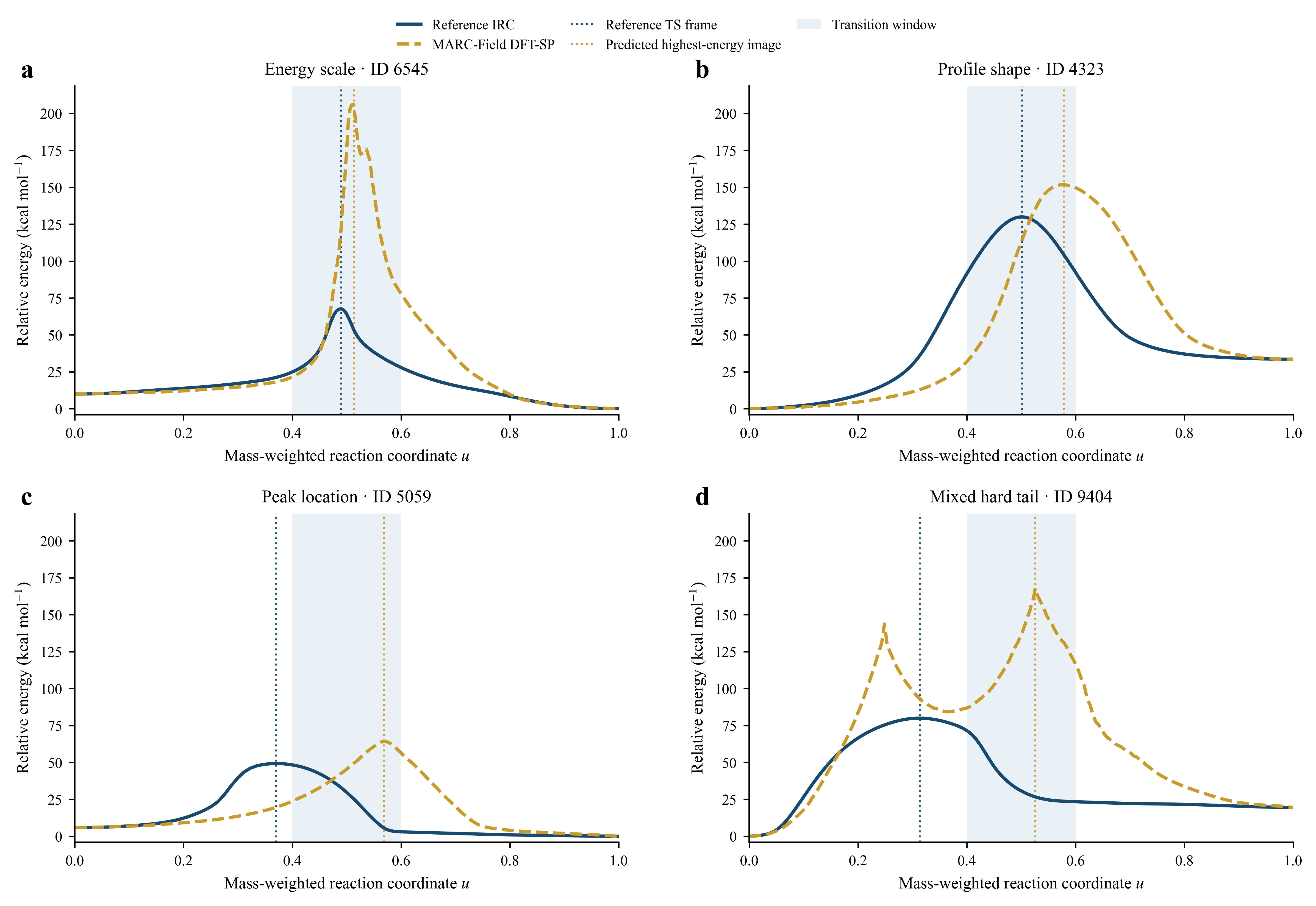}
\par\smallskip
\begin{minipage}{0.98\linewidth}\small
Supplementary Fig. S9 Distinct failure modes in MARC-Field DFT
single-point energy profiles. \textbf{a--d}, Reactions 6545, 4323, 5059
and 9404 illustrate energy-scale, profile-shape, maximum-location and
mixed high-error failures. Solid blue and dashed ochre curves show
reference IRC and MARC-Field profiles. Vertical markers identify the
reference transition-state frame and predicted highest-energy image;
shading marks \emph{u} = 0.40--0.60. Cases were selected to represent
complementary failure modes rather than the four largest values of one
metric.
\end{minipage}
\end{figure}

\begin{figure}[H]
\centering
\includegraphics[width=\linewidth,height=0.62\textheight,keepaspectratio]{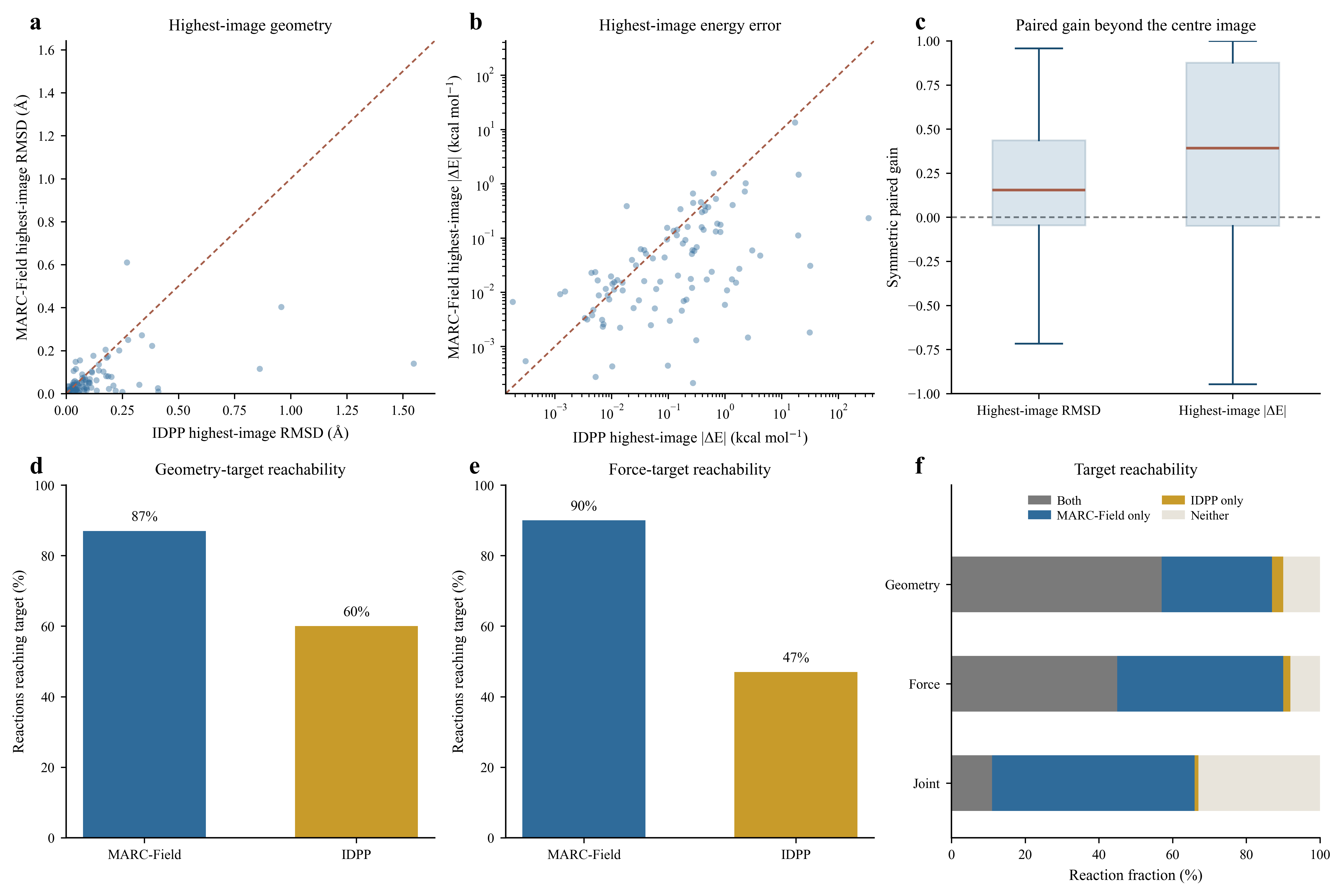}
\par\smallskip
\begin{minipage}{0.98\linewidth}\small
Supplementary Fig. S10 Diagnostics for the paired MARC-Field and IDPP
NEB workflows. \textbf{a,b}, Final highest-energy-image geometry and
absolute relative-energy errors for 100 paired reactions. \textbf{c},
Symmetric paired gain, (\emph{e}\textsubscript{IDPP} -
\emph{e}\textsubscript{MARC})/(\emph{e}\textsubscript{IDPP} +
\emph{e}\textsubscript{MARC}); positive values favour MARC-Field.
\textbf{d,e}, Fractions reaching the predefined geometry and force
targets. \textbf{f}, Decomposition of geometry, force and joint-target
attainment into both methods, MARC-Field only, IDPP only or neither. All
pairs remain in the denominators. The workflows use the same endpoints
and 100 post-construction optimizer steps but differ in the DFT
peak-centring scan and NEB phase schedule, as detailed in Supplementary
Methods.
\end{minipage}
\end{figure}

\begin{figure}[H]
\centering
\includegraphics[width=\linewidth,height=0.62\textheight,keepaspectratio]{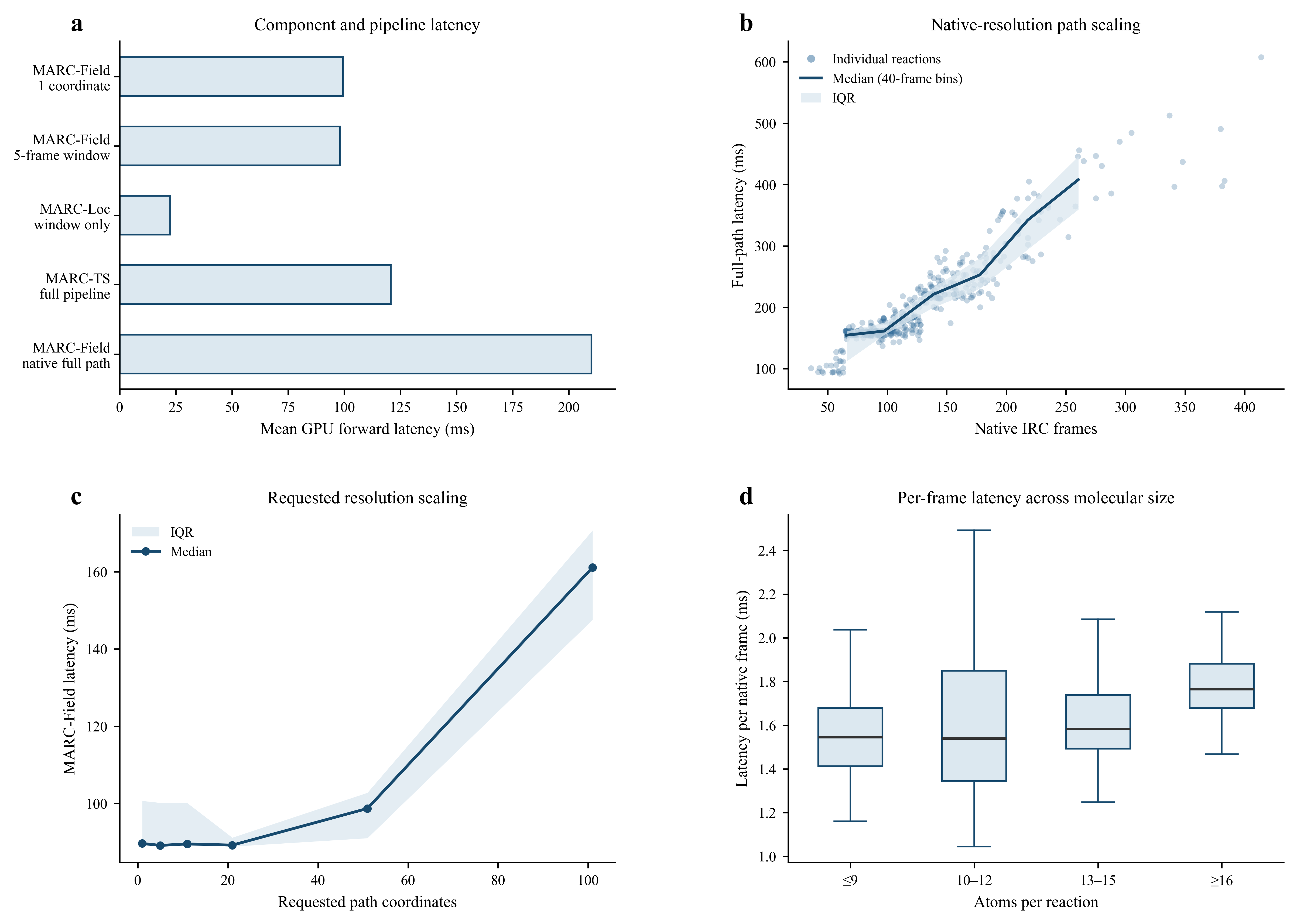}
\par\smallskip
\begin{minipage}{0.98\linewidth}\small
Supplementary Fig. S11 MARC-TS inference latency and resolution scaling.
\textbf{a}, Mean GPU forward latency of model components and the full
MARC-Field-to-MARC-Loc pipeline. \textbf{b}, Native-resolution path
latency versus stored IRC frame count; reactions are grouped in 40-frame
bins and summarized by medians and interquartile ranges. \textbf{c},
MARC-Field latency versus requested coordinate count. \textbf{d},
Per-frame native-path latency across molecular-size bins. Measurements
use FP32 on one NVIDIA RTX A6000 and all 410 held-out reactions.
\end{minipage}
\end{figure}

\clearpage
\section*{Supplementary References}
\begin{enumerate}[leftmargin=*,label={[\arabic*]},itemsep=0.25em,topsep=0.25em]
\small
\item Schreiner, M., Bhowmik, A., Vegge, T., Busk, J. \& Winther, O. Transition1x---a dataset for building generalizable reactive machine learning potentials. Sci. Data 9, 779 (2022).
\item Kabsch, W. A solution for the best rotation to relate two sets of vectors. Acta Crystallogr. A 32, 922--923 (1976).
\item Frisch, M. J. et al. Gaussian 16, Revision C.01 (Gaussian, Inc., 2016).
\item Smidstrup, S., Pedersen, A., Stokbro, K. \& Jønsson, H. Improved initial guess for minimum energy path calculations. J. Chem. Phys. 140, 214106 (2014).
\item Larsen, A. H. et al. The atomic simulation environment---a Python library for working with atoms. J. Phys.: Condens. Matter 29, 273002 (2017).
\item Schütt, K. T., Unke, O. T. \& Gastegger, M. Equivariant message passing for the prediction of tensorial properties and molecular spectra. In Proc. 38th International Conference on Machine Learning 9377--9388 (PMLR, 2021).
\item Duan, C. et al. Optimal transport for generating transition states in chemical reactions. Nat. Mach. Intell. 7, 615--626 (2025).
\item Choi, S. Prediction of transition state structures of gas-phase chemical reactions via machine learning. Nat. Commun. 14, 1168 (2023).
\item Duan, C., Du, Y., Jia, H. \& Kulik, H. J. Accurate transition state generation with an object-aware equivariant elementary reaction diffusion model. Nat. Comput. Sci. 3, 1045--1055 (2023).
\item Galustian, L., Mark, K., Karwounopoulos, J., Kovar, M. P.-P. \& Heid, E. GoFlow: efficient transition state geometry prediction with flow matching and E(3)-equivariant neural networks. Digital Discovery 4, 3492--3501 (2025).
\item Kim, S., Woo, J. \& Kim, W. Y. Diffusion-based generative AI for exploring transition states from 2D molecular graphs. Nat. Commun. 15, 341 (2024).
\end{enumerate}

\end{document}